\documentclass[12pt]{article}

\usepackage{hyperref}

\pdfoutput=1
\usepackage{graphicx}

\usepackage{graphics,psfrag}
\usepackage{amsthm,amssymb,epsfig,amsmath,euscript,array,cite}
\usepackage{color}
\usepackage{graphicx}
\usepackage{amscd}
\usepackage{slashed}

\usepackage[utf8]{inputenc}

\newcommand{\be}{\begin{equation}}
\newcommand{\ee}{\end{equation}}
\def\bea{\begin{eqnarray}}
\def\eea{\end{eqnarray}}

\newcommand\fft[2]{{\frac{#1}{#2}}}

\newcommand\nn{\nonumber}

\def\nn{\nonumber}

\newcommand{\beq}{\begin{equation}}
\newcommand{\eeq}{\end{equation}}
\newcommand{\ben}{\begin{eqnarray}}
\newcommand{\een}{\end{eqnarray}}
\newcommand{\bes}{\begin{subequations}}
\newcommand{\ees}{\end{subequations}}
\newcommand{\blg}{\begin{align}}
\newcommand{\elg}{\end{align}}

\newcommand{\cN}{{\cal N}}
\def\one{\mbox{1 \kern-.59em {\rm l}}}
\def\dphi1{{\dot\phi_1}}
\def\dphi2{{\dot\phi_2}}
\def\dphi3{{\dot\phi_3}}
\def\dphi{{\dot\phi}}

\def\={\, =\, }
 \def\cN{{\cal N}}

\makeatletter \@addtoreset{equation}{section} \makeatother

\begin{document}

\begin{titlepage}

\vspace{14pt}

\hfill LCTP-19-18
\begin{center}

\vspace{1cm}
{\Large \bf Aspects of AdS$_2$ classification in M-theory:}\\

\vspace{.8cm}
{\Large \bf  Solutions with mesonic and baryonic charges}\\

\vspace{.4cm}
%{\large \bf  Toward AdS$_2$ classification with mesonic and baryonic charges in M-theory}\\

%{\Large \bf }\\

\vspace{1cm}

{\bf  Junho Hong$^{a}$, Niall T. Macpherson$^{b}$ and  Leopoldo A. Pando Zayas$^{c}$}

%\vspace{1cm}

{\bf }

\vspace{.4cm}
{\it ${}^{a,c}$  Leinweber Center for Theoretical Physics,  Department of Physics}\\
{\it University of Michigan, Ann Arbor, MI 48109, USA}\\

\vspace{.4cm}
{\it ${}^b$ SISSA International School for Advanced Studies}\\
{\it Via Bonomea 265, 34136 Trieste }\\
{\it and }\\
{\it   INFN, sezione di Trieste}\\
\vspace{.4cm}

{\it ${}^c$ The Abdus Salam International Centre for Theoretical Physics}\\
{\it Strada Costiera 11, 34014 Trieste, Italy}\\

\vspace{.4cm}

\vspace{14pt}

\end{center}
\begin{abstract}
We construct necessary and sufficient geometric conditions for a class of AdS$_2$ solutions of M-theory with, at least, minimal supersymmetry to exist.  We generalize previous results in the literature for ${\cal N}=(2,0)$  supersymmetry in AdS$_2$ to ${\cal N}=(1,0)$. 
When the solution can be  locally described as AdS$_2\times \Sigma_g \times \,$SE$_7$ with $\Sigma_g$ a Riemann surface of genus $g$ and SE$_7$ a seven-dimensional Sasaki-Einstein manifold, we clarify and unify various solutions present in the literature. In the case of SE$_7=Q^{1,1,1}$ we find a new solution with baryonic and mesonic charges turned on simultaneously. 
\end{abstract}

\vspace{1cm}
{\tt junhoh@umich.edu, nmacpher@sissa.it, lpandoz@umich.edu}

\end{titlepage}

%\setcounter{page}{1} \renewcommand{\thefootnote}{\arabic{footnote}}
%\setcounter{footnote}{0}
%\setcounter{page}0
%\newpage

\tableofcontents

%%%%%%%%%%%%%%%%%%%%%%%%%%%%%%%%%%%%%%%%%%%%%%%%%
\section{Introduction}

Our understanding of asymptotically AdS$_4$ black holes has recently undergone a revolutionary transformation. Via the AdS/CFT correspondence the macroscopic entropy of a large class of asymptotically AdS$_4$ black holes has been given a microscopic explanation in terms of states in the dual 3d $\cN=2$ supersymmetric field theories.  Namely, it has been shown that the topologically twisted index in field theory accounts for the macroscopic entropy of magnetically charged asymptotically AdS$_4$ black holes \cite{Benini:2015eyy,Benini:2016rke}. The original observation has been generalized to include various situations: black holes with hyperbolic horizons \cite{Cabo-Bizet:2017jsl}, black holes in massive type IIA supergravity  \cite{Benini:2017oxt,Hosseini:2017fjo}, black holes in universal sectors of gauged supergravities related to M2 branes \cite{Azzurli:2017kxo} and to M5 branes  \cite{Gang:2019uay} (for  reviews with complete lists of references, see \cite{Hosseini:2018qsx,Zaffaroni:2019dhb})\footnote{Very  recently, a microscopic foundation for the entropy of certain rotating, electrically charged,  asymptotically AdS$_4$ black holes has been provided via the superconformal index in  \cite{Choi:2019zpz} and using supersymmetric localization in   \cite{Nian:2019pxj}.}. 

This remarkable progress points to an interesting gap -- {\it there seem to be a number of black hole solutions missing.} Namely, there are various  field theory results for which the dual black holes are not yet known; this is a welcome challenge for the supergravity community. For example, the topologically twisted index can be computed for a fairly generic class of 3d $\cN=2$ supersymmetric field theories, many of which have known gravity duals \cite{Hosseini:2016tor,Hosseini:2016ume}. In particular, there are results for field theory duals to AdS$_4\times\, $SE$_7$ when SE$_7= \{S^7,	Q^{1,1,1}, M^{1,1,1}, V^{5,2}, N^{0,1,0}\}$. The first entry in this list is well understood on the black hole side \cite{Benini:2015eyy,Benini:2016rke}, thanks to known solutions in gauged supergravity \cite{Cacciatori:2009iz} and their eleven-dimensional uplift \cite{Cvetic:1999xp}, which includes the vector multiplets associated to the isometry of $S^7$ and therefore the mesonic charges in the field theory side. For the other entries, the eleven-dimensional uplifts with the corresponding vector and hyper mutiplets are also known \cite{Cassani:2012pj}. In these cases, however, the vector multiplets are associated to the non-trivial two cycles of the internal manifolds and not to isometries of the space.  This implies that the eleven-dimensional uplift with the internal manifolds $Q^{1,1,1}, M^{1,1,1}, V^{5,2}, N^{0,1,0}$ allows only for the AdS$_4$ black holes whose $U(1)$ charges are baryonic in the field theory side, which has been studied in \cite{Halmagyi:2013sla,Halmagyi:2014qza}: the AdS$_4$ black hole solutions with mesonic twists are still missing in this case. Considering that the dual topologically twisted indices with mesonic charges have been already computed in many examples \cite{Hosseini:2016tor,Hosseini:2016ume}, finding such black holes is an interesting and important problem which might generalize the match found in \cite{Benini:2015eyy,Benini:2016rke} to general SE$_7$ internal manifolds.

In this manuscript we do not directly tackle the construction of such missing black holes, rather, we focus on a more modest problem. We focus on understanding the near horizon geometry of those extremal black holes. Such solution must contain an AdS$_2$ region and, in all the  known cases, it turns out that the near horizon region is by itself a solution of the supergravity equations of motion. Thus, the classification of solutions with an AdS$_2$ factor although motivated by understanding extremal black holes in AdS$_4$  is a well-defined problem in supergravity. In this manuscript we take some steps into the full systematic classification. We will, however, be guided by a particularly interesting class of solutions -- AdS$_4\times \,$SE$_7$ -- whose relation to the black hole is presaged  by various field theory computations.

There have been various explicit efforts towards constructing solutions with AdS$_2$ factors, for example,  \cite{Kim:2006qu,Gauntlett:2006ns,Donos:2008ug,Donos:2012sy}. There are other approaches more focused on constructing new solutions and downplaying a classificatory goal such as in \cite{Azzurli:2017kxo}. One of our goals in this manuscript is to take one  systematic step towards a complete classification within a well-defined class.  Another important goal of this paper is to present explicit solutions capable of elucidating some of the current puzzles marring the gravitational understanding of some of the topologically twisted indices on the dual field theory.

The rest of the manuscript is organized as follows. In section \ref{Sec:Class}  we discuss general conditions for the existence of supersymmetric  AdS$_2$ solution in M-theory supporting an SU(4)-structure. We pay particular attention to differences between ${\cal N}=(1,0)$ and ${\cal N}=(2,0)$. Section \ref{Sec:Solns} starts by briefly describing a number of puzzling situations regarding the entropy of asymptotically AdS$_4$ solutions from the AdS/CFT point of view. We then proceed to  cast a number of known solutions within our classificatory scheme and also present the details and motivation for the construction of new solutions with baryonic and mesonic charges turned on.  In appendix \ref{App:NoGoSpin} we demonstrate the absence of AdS$_2$ solutions in M-theory with Spin(7)-structure, while in appendix \ref{App:NoGoAdS3} show that no AdS$_3$ solutions follow from SU(4) structure. We also present some explicit details of the Killing spinor equations in appendix \ref{App:KillingSpinor} to aid readers who prefer a more classic approach to solution building.

%%%%%%%%%%%%%%%%%%%%%%%%%%%%%%%%%
\section{AdS$_2$ solutions with SU(4)-structure}\label{Sec:Class} 
In this section we will derive sufficient geometric conditions for a certain class of minimally supersymmetic  AdS$_2$ solutions in M-theory to exist. The class in question consists of  internal spaces that support an SU(4)-structure. From these conditions we also derive necessary conditions for $\mathcal{N}=(2,0)$ with SU(4)-structure, which turn out to coincide with those presented in \cite{Donos:2008ug}.

%%%%%%%%%%%%%%%%%%%%%%%%%%%%%%%%%
\subsection{Supersymmetric AdS$_2$}
The most general form of a solution to 11 dimensional supergravity that respects the isometries of AdS$_2$ can take a metric and flux that can be decomposed as
\beq\label{eq:AdS2decomp}
ds^2=e^{2A}ds^2(\text{AdS}_2)+ ds^2(\text{M}_9),~~~~~ F= e^{2A}\text{vol}(\text{AdS}_2)\wedge G_2+ G_4,
\eeq
where $e^{2A}$ depends on directions on M$_9$ only, and likewise the 2 and 4 forms $G_2$, $G_4$. We will allow for non trivial $G_4$, so we have both electric and magnetic flux components turned on in general, which should be contrasted with \cite{Kim:2006qu} that only considered the former. In terms of  \eqref{eq:AdS2decomp} the Bianchi identity of the flux, away from possible localised sources, becomes
\beq\label{eq:BI}
d(e^{2A}G_2)=0,~~~~dG_4=0,
\eeq
while its equation of motion is implied by
\begin{subequations}
\begin{align}
&d(e^{2A}\star_9 G_4)+e^{2A}G_2\wedge G_4=0,\label{eq:EOM1}\\[2mm]
&d\star_9 G_2- \frac{1}{2}G_4\wedge G_4=0.\label{eq:EOM2}
\end{align}
\end{subequations}
Of course, for a solution to exist one also needs to solve Einstein's equation, however, as we shall establish later, for the class of supersymmetric solution we are interested in it turns out that these are implied by \eqref{eq:BI} and \eqref{eq:EOM2}.

When supersymmetric, an M-theory solution supports an associated 11 dimensional Killing spinor $\epsilon$ that obeys
\beq\label{eq:11dKSE}
\nabla_N\epsilon+ \frac{1}{4!}\bigg(3F\Gamma_N- \Gamma_N F\bigg)\epsilon=0,
\eeq
where in this expression $F$ should be understood as acting under the Clifford map $F \to \frac{1}{4!}F_{MNPQ}\Gamma^{NMPQ}$. 
For AdS$_2$ solutions we may decompose $\epsilon$ in general as
\beq\label{eq:AdSspinor}
\epsilon= \zeta_+\otimes \chi_1+ \zeta_- \otimes \chi_2,
\eeq
where $\chi_{1,2}$ are two Majorana spinors on M$_9$ and $\zeta_{\pm}$ are Majorana Killing spinors on unit AdS$_2$ of $\pm$ chirality, and so  obey
\beq\label{eq:AdS2KSE}
\nabla_a \zeta_{\pm}= \frac{1}{2} \gamma_a \zeta_{\mp}.
\eeq 
Upon plugging \eqref{eq:AdSspinor} into \eqref{eq:11dKSE} it is then a rather simple matter to show that
\beq\label{eq:chidef}
|\chi|^2=e^A,~~~~ \chi= \frac{1}{\sqrt{2}}(\chi_1+ i \chi_2),
\eeq
where in the first equality we fix an arbitrary constant without loss of generality. One could proceed in kind, using the Killing spinor equations to construct spinor bi-linears and the set of necessary and sufficient geometric conditions they must obey for supersymmetry to be satisfied. However, a more efficient approach is build on an existing geometric classification, namely that of the geometry following from a single arbitrary Killing spinor in 11 dimensions \cite{Gauntlett:2002fz,Gauntlett:2003wb}. This derivation is similar to those in \cite{MacConamhna:2006nb,Katmadas:2015ima}, which look 
at related problems from the same geometric starting point.

\subsection{Review of the Geometry of 11 dimensional Killing spinors}
In \cite{Gauntlett:2002fz,Gauntlett:2003wb} the geometry that follows from a single Majorana Killing spinor $\epsilon$ in M-theory is classified. The fundamental objects of this construction are the 1,2 and 5-forms with components defined in terms on $\epsilon$ as
\begin{align}\label{eq:11dfromsdef}
K_M&=\overline{\epsilon}\Gamma_{M}\epsilon,\\[2mm]
\Xi_{M N}&=\overline{\epsilon}\Gamma_{M N}\epsilon,\\[2mm]
\Sigma_{M NOPQ}&=\overline{\epsilon}\Gamma_{M NOPQ}\epsilon,
\end{align}
respectively, which obey the following algebraic relations
\beq
\iota_K\Xi =0,~~~~\iota_K\Sigma = \frac{1}{2}\Xi\wedge \Xi,
\eeq
among others. Using only the Killing spinor equation \eqref{eq:11dKSE},  the authors of \cite{Gauntlett:2002fz} were able to establish the following necessary and sufficient geometric conditions for supersymmetry
\begin{align}
d\Xi&= \iota_K F,\label{eq:11dsusy1}\\[2mm]
d\Sigma&= \iota_K \star F-\Xi\wedge F,\label{eq:11dsusy2}\\[2mm]
dK&=\frac{2}{3} \iota_{\Xi}F+ \frac{1}{3} \iota_{\Sigma} \star F,\label{eq:11dsusy3}
\end{align}
and that $K$ is a nowhere vanishing Killing vector of both the metric and 4-form flux that can be either time-like or null. The time-like case was considered in \cite{Gauntlett:2002fz} where  \eqref{eq:11dsusy1}-\eqref{eq:11dsusy3} were shown to be sufficient conditions for supersymmetry with the specific analysis performed with respect to an SU(5)-structure (see for instance \cite{Prins:2016zhg} for a review of SU(N) structures)  supported by the internal space orthogonal to $K$. In \cite{Gauntlett:2003wb} the null case was studied, where  \eqref{eq:11dsusy1}-\eqref{eq:11dsusy3} were once more found to be sufficient for supersymmetry giving rise to a $(\text{spin}(7)\ltimes\mathbb{R}^8)\times \mathbb{R}$-structure. In the time-like case supersymmetry together with the the Bianchi identity and equation of motion of the flux implies Einstein’s equations, while in the null case one needs to still solve one component of Einstein’s equations.

In what follows we will be interested in AdS$_2$ solutions that support an SU(4)-structure on their internal space (we also rule out the possibility of Spin(7)-structure in appendix \ref{App:NoGoSpin}), as we shall see, for these $K$ is necessarily time-like. In principle one could work with the SU(5)-structure conditions presented in \cite{Gauntlett:2002fz}, however we find it easier to work directly with \eqref{eq:11dsusy1}-\eqref{eq:11dsusy3}. Before moving on, let us stress that this class of solutions is by no means exhaustive - indeed all AdS$_3$ solutions admit a parametrisation in terms of an AdS$_2$ factor, and we prove in appendix \ref{App:NoGoAdS3} that no such solutions lie in this class.

\subsection{$\mathcal{N}=(1,0)$ AdS$_2$ solutions with symplectic SU(4)-structure}
Our task in this section is to derive $K,\Xi,\Sigma$ under the assumption that a solution decomposes as a warped product of AdS$_2\times\text{M}_9$, with M$_9$ supporting an SU(4)-structure. Then we derive sufficient conditions for a supersymmetric solution in terms of geometric conditions on the SU(4)-strucutre. To this end we need to know the form the bi-linears take on AdS$_2$ and for an SU(4)-structure in 9 dimensions. 

\subsubsection{Bi-linears in 11d in terms of those on AdS$_2$ and M$_9$}
The bi-linears for AdS$_2$ can be derived in general from the Killing spinor equation \eqref{eq:AdS2KSE}, indeed this was already done in \cite{Legramandi:2018qkr} so we will be brief and refer the reader to that reference for further details. We parametrise warped AdS$_2$ as the Poincare patch
\beq\label{eq:warpedads2}
e^{2A}ds^2(\text{AdS}_2)= -(e^0)^2+ (e^r)^2,~~~~~ e^0= e^A r dt,~~~~~e^r= e^A\frac{ dr}{r}, 
\eeq
with $ds^2(\text{AdS}_2)$ of unit radius, and take the real basis of flat space gamma matrices $\sigma_{\mu}$ for $\mu=0,1$ with $\sigma_i$ the Pauli matrices and where $\sigma_{0}=i \sigma_2$. A consequence of this is that $\sigma_3$ is the chirality matrix and so the space-time solutions to \eqref{eq:AdS2KSE} (those giving rise to space-time  supercharges rather than conformal ones) can without loss of generality be taken to be
\beq
\zeta_{+}= \left(\begin{array}{c}\sqrt{r}\\0\end{array}\right),~~~~\zeta_{-}= \left(\begin{array}{c}0\\\sqrt{r}\end{array}\right),
\eeq
which are real and so Majorana in these conventions. We can then easily derive the following 0-2 forms
\begin{subequations}\label{eq:AdSbilinars}
\begin{align}
&\overline{\zeta_{\pm}}\zeta_{\pm}=0,~~~~~ \overline{\zeta_{\pm}}\zeta_{\mp}=\pm r^2,\\[2mm]
&\overline{\zeta_{\pm}}\sigma_{\mu}\zeta_{\pm} dx^{\mu}= r(e^0\pm e^r),~~~~~\overline{\zeta_{\pm}}\sigma_{\mu}\zeta_{\mp}dx^{\mu}=0,\\[2mm]
&\frac{1}{2}\overline{\zeta_{\pm}}\sigma_{\mu}\sigma_{\nu}\zeta_{\pm} dx^{\mu}\wedge dx^{\nu}=0,~~~~~\frac{1}{2}\overline{\zeta_{\pm}}\sigma_{\mu}\sigma_{\nu}\zeta_{\mp} dx^{\mu}\wedge dx^{\nu}= r e^0\wedge e^r,
\end{align}
\end{subequations}
where $\overline{\zeta}= (\sigma_0\zeta)^{\dag}$.

We assume that the internal 9-manifold supports an SU(4)-structure, this means that the 9 dimensional internal spinors $\chi_i$ that appear in \eqref{eq:AdSspinor} should be independent, non zero, and have the same chirality when viewed as spinors in 8 dimensions. An SU(4)-structure has canonical dimension 8, as such  the internal 9-manifold decomposes as
\beq\label{eq:metric}
ds^2(\text{M}_9)=V^2+ds^2(\text{M}_8)
\eeq
where $V$ is a real 1-form and where $\text{M}_8$ supports the SU(4)-structure with associated  $(1,1)$ and $(4,0)$ forms $J$ and $\Omega$ which obey
\beq
\frac{1}{4!}J^4= \frac{1}{2^4}\Omega\wedge \overline{\Omega}= \text{Vol}(\text{M}_{8}),~~~~J\wedge \Omega	=0.
\eeq
The 1-form $V$ lies strictly orthogonal to the SU(4)-structure forms and so
\beq
\iota_V J= \iota_V \Omega=0,~~~~~ \text{Vol}(\text{M}_9)= V\wedge \text{Vol}(\text{M}_8).
\eeq
We need one more piece of information about the forms $(V,J,\Omega)$, exactly how they follow from $\chi_i$. This is given for instance in \cite{Gauntlett:2003cy}, specifically one has
\beq\label{eq:SU4bilinears}
e^{A}V_a=\chi^{\dag}\gamma_a\chi,~~~~~e^{A}J_{a_1a_2}=-i \chi^{\dag}\gamma_{a_1a_2}\chi,~~~~~ e^{A}\Omega_{a_1a_2a_3a_4}= \chi^{c\dag}\gamma_{a_1a_2a_3a_4}\chi,
\eeq 
where $\chi$ is defined in terms of $\chi_i$ as in \eqref{eq:chidef}, $\gamma_a$ are 9 dimensional gamma matrices and the superscript $c$ denotes Majorana conjugation (ie $\chi^c=\frac{1}{\sqrt{2}}(\chi_1-i \chi_2)$ as $\chi_i$ are Majorana). There exists a canonical frame where  $\chi_i$ obey the projections
\beq
\gamma_{1234}\chi_i= \gamma_{5678}\chi_i= \gamma_{1256}\chi_i=-\chi_i
\eeq
and
\beq
\gamma_{1357}\chi_1= + \chi_1,~~~~\gamma_{1357}\chi_2= - \chi_2.
\eeq
The forms are then also canonical, namely
\begin{align}\label{eq:cannonicalforms}
V&= e^9,\\[2mm]
J&= e^{1}\wedge e^{2}+ e^{3}\wedge e^{4}+ e^{5}\wedge e^{6}+ e^{7}\wedge e^{8},\\[2mm]
\Omega&= (e^1+ i e^2)\wedge (e^3+ i e^4)\wedge (e^5+ i e^6)\wedge (e^7+ i e^8),
\end{align}
with $e^{1},...,e^{9}$ a privileged vielbein spanning M$_9$ associated to the canonical frame.

Finally, before we can  put this all together and construct the 11 dimensional forms of \eqref{eq:11dfromsdef} we need to choose a basis of gamma matrices in 11 dimensions consistent with a $2+9$ split and our real gamma matrices on AdS$_2$ -  a natural choice is
\beq\label{eq:11dgammas}
\Gamma_{\mu}= \sigma_{\mu}\otimes\mathbb{I},~~~~~\Gamma_{a}= \sigma_3\otimes \gamma_a,
\eeq
with 9 dimensional intertwiner $B$ such that $\chi^c=B \chi^{*}$, $BB^*=1$ and $B^{-1}\gamma_a B= \gamma_a^*$.
One can then simply insert our spinor \eqref{eq:AdSspinor} and the above gamma matrices into \eqref{eq:11dfromsdef} and using the bilinear expressions \eqref{eq:AdSbilinars} and \eqref{eq:SU4bilinears} establish that
\begin{subequations}
\begin{align}
K&=- e^{A}r e^0,\label{eq:forms1}\\[2mm]
\Xi&=e^{A}r(e^r\wedge V-J),\label{eq:forms2}\\[2mm]
\Sigma&=e^{A}r(- e^0\wedge e^r \wedge V\wedge J+\frac{1}{2}e^0\wedge J\wedge J+ e^r\wedge \text{Re}\Omega+ V\wedge \text{Im}\Omega)\label{eq:forms3}.
\end{align}
\end{subequations}
As $K$ is parallel to $e^0$ we clearly have that $||K||^2=- e^{2A}r^2$, confirming our earlier claim that SU(4)-structure implies a time-like Killing vector in 11 dimensions.

In the next section we present sufficient conditions the SU(4)-structure forms must obey for supersymmetry to hold, and the additional conditions one must satisfy to have a solution.  
%%%%%%%%%%%%%%%%%%%%%%%%%%%%%%
\subsubsection{$\mathcal{N}=(1,0)$ and symplectic SU(4)-structure conditions}\label{eq:neq1ads2}
Given \eqref{eq:forms1}-\eqref{eq:forms3}, and the fact that the solutions we seek respect the isometries of AdS$_2$, it is possible to derive a set of sufficient conditions on  $(e^A,V,J,\Omega)$ only that imply the 11 dimensional conditions \eqref{eq:11dsusy1}-\eqref{eq:11dsusy3}. Upon plugging the former into the later\footnote{We make use of the fact that $\iota_{\Xi}F= -\star(\Xi\wedge \star F)$ and $\iota_{\Sigma}\star F= \star(\Sigma\wedge  F)$.} one finds the following differential conditions 
\begin{subequations}
\begin{align}
&d(e^A J)=0,\label{eq:diffcond1}\\[2mm]
&d(e^{2A}V) + e^{A}J+ e^{2A} G_2=0,\label{eq:diffcond2}\\[2mm]
&d(e^A V\wedge \text{Im}\Omega)-e^{A} J\wedge G_4=0,\label{eq:diffcond3}\\[2mm]
&\star_9\big(2V\wedge \star_9 G_2+  \text{Re}\Omega\wedge G_4\big)+6 dA=0,\label{eq:diffcond4}\\[2mm]
&d(e^{2A} \text{Re}\Omega)- e^{A} V\wedge \text{Im}\Omega+ e^{2A}(\star_9 G_4-V\wedge G_4)=0,\label{eq:diffcond5}
\end{align}
\end{subequations}
and the algebraic constraints
\begin{subequations}
\begin{align}
&J\wedge J\wedge G_4=0,\label{eq:algebcond1}\\[2mm]
&V\wedge (\text{Im}\Omega \wedge G_2+ J\wedge G_4)=0,\label{eq:algebcond2}\\[2mm]
&e^{A}\big(2J\wedge \star_9 G_2-  V\wedge \text{Im}\Omega\wedge G_4\big)=6 \text{Vol}_9\label{eq:algebcond3},
\end{align}
\end{subequations}
where we have simplified expressions wherever possible and omitted conditions that are obviously implied by others. It seem likely that these are necessary and sufficient conditions for supersymmetry - sufficiency is guaranteed as \eqref{eq:11dsusy1}-\eqref{eq:11dsusy3} are themselves necessary and sufficient but we have not totally ruled out the possibility of some redundancy in \eqref{eq:diffcond1}-\eqref{eq:algebcond3}. At any rate this does not overly concern us 
 as solving all of the above guarantees supersymmetry, be there redundancies or not.

Let us now study these conditions and see what we can learn: From \eqref{eq:diffcond1} we see that $e^{-A}ds^2(\text{M}_8)$ in general supports a symplectic SU(4)-structure \cite{Prins:2016zhg}. The electric component of the flux $G_2$, is defined by \eqref{eq:diffcond2} and automatically solves it's Bianchi identity \eqref{eq:BI} due to \eqref{eq:diffcond1}. If one  assumes the Bianchi identity of $G_4$ it is possible to also derive the equation of motion for $G_2$ \eqref{eq:EOM1} by substituting \eqref{eq:diffcond2} and \eqref{eq:diffcond3} into $d$\eqref{eq:diffcond5}.  As such, solving \eqref{eq:diffcond1}-\eqref{eq:algebcond3} is sufficient for supersymmetry and by the integrablity argument of \cite{Gauntlett:2002fz}, one need only additionally impose
\begin{subequations}
\begin{align}
	0&=dG_4,\label{eq:Bianchi}\\
	0&=d(e^{2A}\star_9 G_4)+e^{2A}G_2\wedge G_4,\label{eq:eom}
\end{align}
\end{subequations}
in the absence of localised sources to have a solution\footnote{In the presence of sources the Bianchi identity of $G_4$ will be modified by a localised source term, schematically  $dG_4 = \delta$. The existence of a supersymmetric solution will then also require that the source is calibrated.}.

Additionally we note that if we further decompose 
\beq
G_4=V\wedge G_3+ \hat{G_4}
\eeq
then \eqref{eq:diffcond5} completely determines $G_3$ but does not uniquely fix $\hat{G}_4$, indeed it only fixes $\hat{G_4}-\star_8 \hat{G_4}$ and in particular yields no information about the self dual components of $\hat G_4$, which are only constrained by the remaining conditions.  Lastly, we can exploit \eqref{eq:diffcond4} to derive another condition, namely
\beq
\mathcal{L}_V A=- \frac{1}{6}\text{Re}\Omega {\big\lrcorner}\hat{G}_4
\eeq
which implies that $V$ is a Killing vector with respect to $A$ iff the RHS is zero, but is not so in general.  Indeed by studying the 9 dimensional bi-spinor relations that follow from \eqref{eq:11dKSE} and \eqref{eq:AdSspinor} directly one can show that $V$ is not in general a Killing vector of the internal metric either. This should be contrasted with \cite{Kim:2006qu} which found that AdS$_2$ solutions with only non trivial electric flux turned on always come with a U(1) Killing vector dual to $V$. They further establish that regularity for such solutions is only possible when the internal space supports an SU(4)-structure, which in the end is refined to a Kahler-structure. Although it is not stressed in \cite{Kim:2006qu}, the U(1) in this case is actually an R-symmetry indicating an enhancement of supersymmetry to $\mathcal{N}=(2,0)$, a fact we will confirm in the next section.  However in the presence of magentic flux, this  U(1) is not generically present, so there is no enhancement of supersymmetry. Further the generic structure is symplectic, not Kahler.

In the next section we will derive the sufficient conditions for enhancement to $\mathcal{N}=(2,0)$ supersymetry, as we shall see, this requires $G_4$ to be highly constrained, but does not require it to vanish.

\subsection{$\mathcal{N}=(2,0)$ and Kahler-structure conditions}\label{N=(2,0)}
For $\mathcal{N}=(2,0)$ supersymmety we must modify the spinor ansatz of \eqref{eq:AdSspinor} as
\beq\label{eq:spinor11dneq2}
\epsilon= \sum_{a=1}^2\bigg(\zeta^a_+ \otimes \chi^a_1+ \zeta^a_- \otimes \chi^a_2\bigg),
\eeq
with $\zeta^a_{\pm}$  doublets of independent real spinors on AdS$_2$ of $\pm$ chirality. The 2d $\mathcal{N}=(2,0)$ superconformal algebra  contains a SO(2) R-symmetry under which the internal spinor $\chi^a_1$ and $\chi^a_2$ should be charged - this can be phrased in terms of the spinoral Lie derivative which acts on a spinor $\chi$ along a Killing vector $k$ as
\beq\label{eq:spinorialLiederivative}
\mathcal{L}_k \chi= k^a\nabla_a\chi+ \frac{1}{4}\nabla_a k_b\gamma^{ab}\chi.
\eeq 
The condition that the spinors are charged is then equivalent to imposing that
\beq\label{eq:SO2cond}
\mathcal{L}_{\tilde{V}} \chi^a_{1,2}= -\frac{n}{2}\epsilon^a_{~b} ~\chi^b_{1,2}.
\eeq
where $\tilde{V}$ is an SO(2) Killing vector and $n$ is a non zero integer. One can introduce a local coordinate $\psi$ such that  $\tilde{V}=\partial_{\psi}$ and work in a frame in which the vielbein depends on $\psi$ only in the combination
\beq
\tilde{V}= e^{C}(d\psi+ \rho),
\eeq
with $\mathcal{L}_{\tilde{V}}\rho=0$ making M$_9$ a U(1) fibration over an 8d base that is independent of $\psi$. In such a frame  \eqref{eq:spinorialLiederivative} becomes $\mathcal{L}_{\tilde{V}}\chi=\partial_{\psi}\chi$, making \eqref{eq:SO2cond} relatively easy to solve - namely one should parametrise 
\beq\label{eq:doublets}
\chi^a_1= \left(\begin{array}{c}\cos(\frac{n\psi}{2})\chi^0_1-\sin(\frac{n\psi}{2})\chi^0_2\\ \sin(\frac{n\psi}{2})\chi^0_1+\cos(\frac{n\psi}{2})\chi^0_2\end{array}\right)^a,~~~\chi^a_2= \left(\begin{array}{c}\sin(\frac{n\psi}{2})\chi^0_1+\cos(\frac{n\psi}{2})\chi^0_2\\ -\cos(\frac{n\psi}{2})\chi^0_1+\sin(\frac{n\psi}{2})\chi^0_2\end{array}\right)^a,
\eeq
where $\chi^0_{1,2}$ are orthogonal, independent of $\psi$, and define an SU(4)-structure as before. Notice that \eqref{eq:spinor11dneq2} is the sum of two independent $\mathcal{N}=1$ sub-sectors  parameterised by the spinors that involve $\zeta^1_{\pm}$  and $\zeta^2_{\pm}$ respectively. The form of \eqref{eq:doublets} then ensures that one sub-sector is mapped into the other by sending $\psi\to \psi+ \pi$ and as such, whenever the physical fields of a solution respect the SO(2) isometry, it is sufficient to solve for one of these  $\mathcal{N}=1$ sub-sectors as the other follows automatically enhancing supersymmetry to $\mathcal{N}=(2,0)$. We take 
\beq
\chi_1= \cos(\frac{n\psi}{2})\chi^0_1-\sin(\frac{n\psi}{2})\chi^0_2,~~~~\chi_2=\sin(\frac{n\psi}{2})\chi^0_1+\cos(\frac{n\psi}{2})\chi^0_2,
\eeq
to be this sub-sector, and construct
\beq
\chi= \chi_1+ i \chi_2= e^{\frac{i n}{2}\psi}(\chi^0_1+ i \chi^0_2).
\eeq
It should then be clear from \eqref{eq:SU4bilinears} that the only place $\psi$ enters will be in $\Omega$ and $V$ in the form
\beq\label{eq:psiappearsin}
\Omega= e^{i n \psi}\Omega_0,~~~~\tilde{V}=V= e^{C}(d\psi+ \rho),
\eeq
with $\Omega_0$ defined as in \eqref{eq:SU4bilinears} but for $\chi^0=\chi_1^0+ i\chi^0_2$,
leaving \eqref{eq:diffcond1}- \eqref{eq:algebcond3} otherwise unchanged. Generally one might wonder if the isometry direction could lie partially along $V$ and partially along a 1-form in M$_8$, but the 1-form dual to the Killing vector can only be a linear combination of the 1-form bi-linears we can construct from $\{\chi^1_1~,\chi^2_1,~\chi^1_2,~\chi^2_2\}$ which all yield either 0  or something parallel to V - as such we can safely take $V$ as the 1-form dual to $\partial_{\psi}$. 

We have at this stage imposed that we have an SU(4)-structure containing a U(1) R-symmetry, but to ensure that this is an isometry of the full solution we need to further impose that
\beq
\mathcal{L}_V A=0,~~~~\mathcal{L}_VG_{4}=0,
\eeq
we also require $\mathcal{L}_VG_{2}=0$, but this follows automatically. Imposing these conditions significantly refines \eqref{eq:diffcond1}-\eqref{eq:algebcond3}, for instance given \eqref{eq:psiappearsin}
 the latter means that the parts of \eqref{eq:diffcond5} that involve $G_4$ must vanish by themselves, making $G_4$ self dual on M$_8$. To proceed it is useful to decompose the exterior derivative as
\beq
d= d\psi \partial_{\psi}+ d_8,
\eeq
After some massaging it is then possible to show that  supersymmmetry is implied by the following conditions 
\begin{subequations}
\begin{align}
&d_8(e^A J)=0,\label{eq:BPSneq21}\\[2mm]
&e^C+n e^A=0\label{eq:BPSneq22},\\[2mm]
%&e^{C}d_8\rho\wedge \text{Im}\Omega-J\wedge G_4=0,\label{eq:BPSneq23}\\[2mm]
&d_8(e^{2A}\Omega)-i n e^{2A}\rho\wedge \Omega=0,\label{eq:BPSneq24}\\[2mm]
&e^{A+C}J\wedge \star_8 d_8 \rho+ \text{Vol}_8=0,\label{eq:BPSneq25}\\[2mm]
%&G_4\wedge J\wedge J
&\iota_VG_4=G_4\wedge J=G_4\wedge \Omega= G_4-\star_8 G_4=0,\label{eq:BPSneq26}\\[2mm]
&e^{2A}G_2+ e^{2A+C}d_8\rho+ e^{A}J- e^{-C}V\wedge d_8(e^{2A+C})=0\label{eq:BPSneq27},
\end{align}\label{eq:BPS}
\end{subequations}
with all else that follows from \eqref{eq:diffcond1}-\eqref{eq:algebcond3} implied - one still needs to impose \eqref{eq:Bianchi}-\eqref{eq:eom},  to have a solution. Together \eqref{eq:BPSneq21} and \eqref{eq:BPSneq24} define a warped Kahler-structure with $e^{-A}ds^2(\text{M}_8)$  a Kahler manifold, so we have reproduced the result of \cite{Donos:2008ug}. The condition \eqref{eq:BPSneq26} highly constrains $G_4$, but does not fix it completely. The first condition tells us that $G_4$ is defined on M$_8$ only, then of the 70 functions an arbitary 4-form in 8 dimensions can contain, only 20 are independent once the rest of \eqref{eq:BPSneq26} is imposed. From a practical perspective though, the symmetries of a given solution will fix many more of these terms - in fact in the cannonical frame of \eqref{eq:cannonicalforms} all solutions we are aware of have magnetic flux of the form
\beq
G_4 = f_1(e^{1234}+e^{5678})+ f_2(e^{1256}+e^{3478})+ f_3(e^{1278}+e^{3456}),~~~~ f_1+f_2+f_3=0,
\eeq
with $f_i$ functions with support on M$_8$ only.

%%%%%%%%%%%%%%%%%%%%%%%%%%%%%%%%%%%%%%%%%%%
\section{Toward AdS$_2 \times \Sigma_g \times Q^{1,1,1}$ solutions with baryonic and mesonic charges }\label{Sec:Solns}
%%%%%
In this section we construct explicit solutions with various properties and explain their connection with known solutions in the literature. Before diving into such technical constructions we briefly review the status of such solutions in the context of the AdS/CFT correspondence. 

\subsection{Preamble from  AdS/CFT}

Following the seminal work of ABJM  \cite{Aharony:2008ug} in establishing the now prototypical dual pair of AdS$_4\times S^7/\mathbb{Z}_k$/CFT$_3$,  a plethora of examples was  given.  The gravity side of some of the well established cases are  Freund-Rubin type solutions of the form  AdS$_4\times\, $SE$_7$ for a certain list of seven-dimensional Sasaki-Einstein spaces, SE$_7$.  Some prominent cases in the list  include  SE$_7= \{S^7,	Q^{1,1,1}, M^{1,1,1}, V^{5,2}, N^{0,1,0}\}$. For example, for  $M^{1,1,1}$  which is geometrically  a $U(1)$ bundle over $\mathbb{CP}^2\times S^2$ , the dual quiver Chern-Simons matter theory was discussed in \cite{Martelli:2008si,Hanany:2008cd};  for $Q^{1,1,1}$ which is geometrically a  $U(1)$ bundle over $S^2\times S^2 \times S^2$, the dual theory is an ${\cal N}=2$ supersymmetric Chern-Simons quiver gauge theory with Chern-Simons levels $(k,k,-k,-k)$, see   \cite{Franco:2009sp,Benini:2009qs,Cheon:2011vi}. The non-toric case in the list  AdS$_4 \times  V^{5,2}$ was addressed in \cite{Martelli:2009ga,Cheon:2011vi}. For all these dual pairs the free energy of the field theory on S$^3$ was shown to agree with the regularized on-shell action on the gravity side largely using techniques presented in \cite{Herzog:2010hf} (see also  \cite{Amariti:2019pky} for recent applications). More recently, the topologically twisted index of a number of these field theories has been computed \cite{Hosseini:2016tor,Hosseini:2016ume,Jain:2019lqb,Jain:2019euv}. Given the impressive match of the free energy on S$^3$, it is natural to expect that the topologically twisted index in these cases would be related to the entropy of the dual magnetically charged asymptotically AdS$_4$ black holes. 

A rigorous way to relate the topologically twisted index to the dual magnetically charged AdS$_4$ black holes has been developed recently: the entropy of a class of magnetically charged asymptotically AdS$_4$ black holes can be obtained by \emph{extremizing} the topologically twisted index with respect to the chemical potentials associated to flavor symmetries in the dual 3d ${\cal N}=2$ Chern-Simons matter theory (${\cal I}$-extremization) \cite{Benini:2015eyy,Benini:2016rke,Cabo-Bizet:2017jsl,Benini:2017oxt,Hosseini:2017fjo,Azzurli:2017kxo,Gang:2019uay}. Various groups have recently considered the dual to ${\cal I}$-extremization by studying a class of theories obtained by a twisted compactification of M2-branes living at the tip of a Calabi-Yau fourfold \cite{Gauntlett:2019roi,Hosseini:2019ddy,Kim:2019umc}. There are, however, a number of puzzling facts regarding the allowed space of charges. For example, Hosseini and Zaffaroni showed that the two extremization principles are equivalent for theories without baryonic symmetries \cite{Hosseini:2019ddy}, while  Kim and Kim \cite{Kim:2019umc} considered cases with either only baryonic or only  mesonic fluxes turned on.  Recall that according to the AdS/CFT dictionary mesonic flavor symmetries manifest themselves as isometries in the gravity solution while baryonic symmetries manifest themselves as cohomology cycles on which one can basically wrap M2 or M5 branes. 

The duality between asymptotically AdS$_4$ black holes with general SE$_7$ internal manifolds and their corresponding 3d $\mathcal N=2$ supersymmetric field theories, however, has not yet been understood well enough in cases when the gravity solutions is equipped with general charges dual, to generic flavor charges in the field theory - including mesonic and baryonic charge.  For mesonic symmetries, this is mainly due to the lack of explicit AdS$_4$ black hole solutions and their AdS$_2$ near horizon geometries as we mentioned in the introduction. For baryonic symmetries, the AdS$_2$ near horizon solution has been obtained in certain cases already in \cite{Donos:2008ug,Azzurli:2017kxo} but the issue is that the entropy computed from the near horizon solution does not match the index of the purported dual field theory \cite{Azzurli:2017kxo}. Summarizing, the comparison between the black hole entropy based on explicit AdS$_2$ near horizon solutions  and the supersymmetric topologically twisted index of the dual field theory has not yet been successful for generic SE$_7$ internal manifolds  and generic flavor symmetries.

Constructing solutions with baryonic and mesonic charges for an abstract form of seven-dimensional Sasaki-Einstein manifolds is quite involved and we defer such a treatment for future work. In this section, we focus on $Q^{1,1,1}$ which naturally allows for baryonic charges by virtue of its second Betti number being, $b_2(Q^{1,1,1})=2$ \cite{Herzog:2000rz}. This specifies our goal as to find a $\mathcal N=(2,0)$ AdS$_2$ solution, which corresponds to the near horizon geometry of an AdS$_4$ black hole with the 7-dimensional internal manifold $Q^{1,1,1}$ that is holographically dual to the 3d $\mathcal N=2$ flavored ABJM theory.  In particular, we are interested in the case with non-trivial mesonic charges, since the solution with purely baryonic charges have been already studied in the literature, see \cite{Donos:2008ug,Halmagyi:2013sla,Azzurli:2017kxo} for examples. As we will see below, the road to this solution is winding and we might, on occasions, find branches that take us off the main pathway.

Before getting into technical details, let us first make a few general remarks regarding the general class of black holes in AdS$_4\times SE_7$ and its near horizon geometry  AdS$_2$ region.  Given the near horizon AdS$_2$ solution, for example, we can compute important physical quantities of the corresponding AdS$_4$ black hole such as its  Bekenstein-Hawking entropy. Then the black hole entropy can be related to the supersymmetric index of the dual  $\mathcal N=2$ supersymmetric field theory.  For example, an AdS$_2\times\Sigma_2\times SE_7$ solution equipped with a universal twist is  already known, and the black hole entropy computed from the solution matches the supersymmetric index of the dual $\mathcal N=2$ supersymmetric field theory. In fact, the corresponding full AdS$_4$ black hole solution has been already found and related to the corresponding dual field theories in this case \cite{Hosseini:2016ume,Azzurli:2017kxo}. One goal of the explicit construction we present in this section is to go beyond the universal twist by involving extra flavor symmetries, in particular the mesonic ones.

%%%%%
\subsection{General Ansatz}
\label{general:ansatz}
%%%%%
Following the general results in the previous section, we consider the following metric Ansatz for a $\mathcal N=(2,0)$ supersymmetric  AdS$_2$ background of our interest:
\begin{align}
	ds^2=&~e^{2A(x_1)}\left(-r^2dt^2+\fft{dr^2}{r^2}\right)+e^{2C(x_1)}\left(d\psi+\rho\right)^2\nn\\&+e^{-A(x_1)}\left(f(x_1)\left(\fft{dx^2}{1-kx^2}+(1-kx^2)d\phi^2\right)+f_1\left(\fft{dx_1^2}{U(x_1)}+U(x_1)(d\phi_1+n_1xd\phi)^2\right)\right.\nn\\&\kern5em\left.+\Sigma_{i=2}^3f_i\left(\fft{dx_i^2}{1-x_i^2}+(1-x_i^2)d\phi_i^2\right)\right),\label{metric}
\end{align}
where $\rho$ is defined as
\begin{equation}
	\rho=nxd\phi+g_1(x_1)(d\phi_1+n_1xd\phi)+g_2x_2d\phi_2+g_3x_3d\phi_3,
\end{equation}
and $k=1,0,-1$ for the Riemann surface with real coordinates $(x,\phi)$ being $S^2,T^2,H^2$ respectively. We define the natural co-frame as
\begin{align}
	&e^1=e^{A(x_1)}rdt,\quad e^2=e^{A(x_1)}\fft{dr}{r},\quad e^3=e^{C(x_1)}(d\psi+\rho),\nn\\
	&e^4=e^{-A(x_1)/2}\sqrt{f(x_1)}\fft{dx}{\sqrt{1-kx^2}},\quad e^5=e^{-A(x_1)/2}\sqrt{f(x_1)}\sqrt{1-kx^2}d\phi,\nn\\& e^6=e^{-A(x_1)/2}\sqrt{f_1}\fft{dx_1}{\sqrt{U(x_1)}},\quad e^7=e^{-A(x_1)/2}\sqrt{f_1}\sqrt{U(x_1)}(d\phi_1+n_1xd\phi),\nn\\
	&e^{2i+4}=e^{-A(x_1)/2}\sqrt{f_i}\fft{dx_1}{\sqrt{1-x_i^2}},\quad
	e^{2i+5}=e^{-A(x_1)/2}\sqrt{f_i}\sqrt{1-x_i^2}d\phi_i.\quad (i=2,3)\label{coframe}
\end{align}
In terms of this co-frame, the 4-form Ansatz given in equation  (\ref{eq:AdS2decomp}) takes the form:
\begin{subequations}
\begin{align}
	F&=e^1\wedge e^2\wedge G_2+G_4,\\
	G_4&=e^4\wedge e^5\wedge(\Sigma_{i=1}^3L_ie^{2i+4}\wedge e^{2i+5})+\fft12\Sigma_{i\neq j}^3M_{ij}e^{2i+4}\wedge e^{2i+5}\wedge e^{2j+4}\wedge e^{2j+5},
\end{align}\label{4form}
\end{subequations}
where $M_{ij}$ is symmetric on $ij$. 

Before solving the $\mathcal N=(2,0)$ supersymmetric conditions (\ref{eq:BPS}), the 4-form Bianchi identity (\ref{eq:Bianchi}), and the 4-form equation of motion (\ref{eq:eom}) for the above Ansatz (\ref{metric} -- \ref{4form}),  let us explain the geometric meaning of various parameters and functions introduced in the above Ansatz. First, the free parameters $f_1,f_2,f_3$ and $L_i,M_{ij}$ yield non-trivial dyonic charges associated to the two Betti multiplets dual to the charges under the baryonic symmetries in the field theory side \cite{Donos:2008ug,Azzurli:2017kxo}. Recall that there are exactly two Betti multiplets in this case since $Q^{1,1,1}$ has two non-trivial 2-cycles. Second, a parameter $n_1$ added along one of the $U(1)$ isometries of $Q^{1,1,1}$ as $d\phi_1\to d\phi_1+n_1xd\phi$ is expected to be dual to a mesonic charge in the field theory side. The functions  $A(x_1),f(x_1),U(x_1)$ are introduced to allow for the deformation of the Ansatz by this parameter $n_1$. Note that a similar deformation was attempted in  \cite{Azzurli:2017kxo} although without any success for non-vanishing $n_1$.

Now we return to solving  (\ref{eq:BPS}), (\ref{eq:Bianchi}), and (\ref{eq:eom})  with the above Ansatz (\ref{metric} -- \ref{4form}). Let us first identify the differential forms defining the SU(4)-structure and the orthogonal one-form $V$:
\begin{subequations}
\begin{align}
	V&=e^3,\\
	J&=\sum_{i=0}^3e^{2i+4}\wedge e^{2i+5},\\
	\Omega&=e^{im\psi}(e^4+ie^5)\wedge (e^6+ie^7)\wedge (e^8+ie^9)\wedge (e^{10}+ie^{11}).
\end{align}\label{SU(4)}
\end{subequations}
Under the above identification (\ref{SU(4)}), the 2-form  $G_2$ is determined by one of the supersymmetry conditions, namely, (\ref{eq:BPSneq27}). 
Then the remaining $\mathcal N=(2,0)$ supersymmetry conditions in (\ref{eq:BPS}) yield
\begin{subequations}
\begin{align}
	e^{C(x_1)}&=-m e^{A(x_1)},\\
	m&=k/n=1/g_2=1/g_3,\\
	f(x_1)&=f_0+f_1n_1x_1,\label{Susy:f}\\
	g_1(x_1)&=-\fft{f_1n_1U(x_1)+f(x_1)U'(x_1)}{2mf(x_1)},\\
	e^{3A(x_1)}&=\fft{2f_1f_2f_3f(x_1)}{2f_1f_2f_3(k-n_1U'(x_1))+f(x_1)(2f_1(f_2+f_3)-f_2f_3U''(x_1))},\label{Susy:eA}\\
	\Sigma_iL_i&=0,\\
	M_{ij}&=|\epsilon_{ijk}|L_k.
\end{align}\label{geometric:susy}
\end{subequations}
Next, the 4-form Bianchi identity (\ref{eq:Bianchi}) yields
\begin{subequations}
\begin{align}
	L_1&=c_1e^{2A(x_1)},\\
	L_2&=\Big(-\fft{c_1}{2}+\fft{c_2}{f(x_1)^2}\Big)e^{2A(x_1)},
\end{align}\label{geometric:Bianchi}
\end{subequations}
where $c_1$ and $c_2$ are arbitrary constants. Finally, the 4-form equation of motion (\ref{eq:eom}) reduces to the following 4-th order non-linear ordinary differential equation (ODE) for $U(x_1)$:
\begin{align}
	0=&~f_1^2f_2f_3(4c_2^2+f_1^2n_1^4U(x_1)^2)-2f(x_1)f_1^3f_2f_3n_1^3\left(U(x_1)U'(x_1)\right)\nn\\&+f(x_1)^2f_1^2f_2f_3n_1^2\left(U'(x_1)^2+U(x_1)U''(x_1)\right)\nn\\&+f(x_1)^3f_1\left(4f_1(f_2+f_3)(-k+n_1U'(x_1))+f_2f_3(2kU''(x_1)-3n_1(U(x_1)U''(x_1))')\right)\nn\\&+f(x_1)^4\left(-4f_1^2+3f_1^2f_2f_3c_1^2+2f_1(f_2+f_3)U''(x_1)-f_2f_3(U(x_1)U'''(x_1))'\right).\label{geometric:eom}
\end{align}
We have not been able to find the most general solution to this ODE. In the following subsections, we focus on some particular  solutions for $U(x_1)$. We will first recast some solutions already known in the literature into our framework and then discuss some new solutions with various level of interests from the dual field theory point of view. 

Based on the above general Ansatz, we now reproduce previously known solutions with $n_1=0$ in Section \ref{known:sol}.  We search for new solutions of interest with $n_1\neq0$ in Section \ref{new:sol}. It is worth emphasizing that in both analytic and numerical approaches looking for AdS$_2$ solutions with mesonic charges ($n_1\neq0$), the resulting solutions we find are equipped with non-trivial baryonic charges. These baryonic charges cannot be arbitrarily turned off, since they are required for the solutions to be globally well-defined. The following table summarizes what we have briefly discussed here and we will discuss further details below.
\begin{center}
\begin{tabular}{|c|c|c|}
	\hline
	\textbf{Known} & Without mesonic charges & Without baryonic charges (section \ref{universal:sol})\\ \cline{3-3}
	\textbf{solutions} & ($n_1=0$, section \ref{known:sol}) & With baryonic charges (section \ref{Betti:sol}) \\ \hline
	\textbf{New} & With mesonic charges & Analytic approach (section \ref{poly:sol}) \\ \cline{3-3}
	\textbf{solutions} & ($n_1\neq0$, section \ref{new:sol}) & Numerical approach (section \ref{numeric:sol}) \\ \hline
\end{tabular}
\end{center}
%

%%%%%
\subsection{Previously known solutions}\label{known:sol}
%%%%%
%%%%%
\subsubsection{The universal twist solution: AdS$_2\times\Sigma_2\times Q^{1,1,1}$ }\label{universal:sol}
%%%%%

First, we consider a special solution $U(x_1)=1-x_1^2$ to equation (\ref{geometric:eom}), corresponding  to 
\begin{equation}
	n_1=0,\qquad f_1=f_2=f_3,\qquad c_1=c_2,.
\end{equation}
In this case equations (\ref{geometric:susy}), (\ref{geometric:Bianchi}), and (\ref{geometric:eom}) are simplified as
\begin{subequations}
\begin{align}
	e^{C(x_1)}&=-me^{A(x_1)},\\
	m&=k/n=1/g_2=1/g_3,\\
	f(x_1)&=f_0=-kf_1,\\
	g_1(x_1)&=x_1/m,\\
	e^{3A(x_1)}&=f_1/2,\\
	L_i&=M_{ij}=0,
\end{align}\label{universal:twist}
\end{subequations}
Note that the condition $f_0=-kf_1$ comes  from the 4-form equation of motion (\ref{geometric:eom}). Since $f(x_1)=f_0$ and $f_1$ must have the same sign for the metric (\ref{metric}) to be positive definite, $k$ must be chosen as $k=-1$. That is, the solution exists only for a negatively curved Riemann surface $\Sigma_g$ with $g\ge 2$.

Under the coordinate transformation $z=1/r$ and the following identifications,
\begin{align}
	f_1\to \fft{L^3}{32},\quad m\to 4,\quad kxd\phi\to A,\quad \fft14\sum_{i=1}^3x_id\phi_i\to\sigma,
\end{align}
the solution (\ref{metric},\,\ref{4form}) with (\ref{universal:twist}) can be rewritten as
\begin{subequations}
\begin{align}
	ds^2&=\fft{L^2}{4}ds_4^2+L^2\left(\left(d\psi+\sigma+\fft14A\right)^2+\fft18\sum_{i=1}^3\left(\fft{dx_i^2}{1-x_i^2}+(1-x_i^2)d\phi_i^2\right)\right), \label{Eq:n10-metric}\\
	ds_4^2&=\fft14\left(\fft{-dt^2+dz^2}{z^2}\right)+\fft12\left(\fft{dx^2}{1+x^2}+(1+x^2)d\phi^2\right),\\
	F&=L^3\left(\fft38\mathrm{vol}_4-\fft18(*_4dA)\wedge d\sigma\right),
\end{align}
\end{subequations}
which is equivalent to the AdS$_2\times\Sigma_2\times Q^{1,1,1}$ solution with a universal twist recently discussed in \cite{Azzurli:2017kxo}. More precisely, note  the presence of the graviphoton in the U(1)  fiber as defined with the co-frame $d\psi +\sigma +\frac{1}{4}A$ above in equation \eqref{Eq:n10-metric}. 

There is a very natural way to interpret this solution geometrically.  Given the form of the U(1) fiber in  Eq. (\ref{Eq:n10-metric}) we can interpret the would be magnetic charge as describing a U(1) bundle over an eight-dimensional space which is Kahler-Einstein, in other words, the metric part of this solution can be interpreted as  AdS$_2 \times $ SE$_9$. With such a metric Ansatz, there is no natural 4-form as would have been the case for the  original Freund-Rubin AdS$_4\times Q^{1,1,1}$. Therefore,  one considers  roughly vol(AdS$_2$)$\wedge$ vol$(\Sigma_g$)  which is  inherited from the vol(AdS$_4$) part and adds vol(AdS$_2$)$\wedge J_{Kahler}$ which is the other natural 4-form given the symmetries; note that this $J_{Kahler}$ is still the one in the base of the SE$_7$.

%%%%%
\subsubsection{Deformed AdS$_2\times\Sigma_g\times Q^{1,1,1}$ solution with Betti multiplets}\label{Betti:sol}

In this subsection we discuss solutions with only the Betti multiplets turned on. They correspond, on the field theory to field theory configurations with baryonic charges turned on. We will start with the general situation and consider its simplification later. 

%%%%%%%%%%%%%%%%%%%%%%%%%%%%
\subsubsection*{Two Betti multiplets}
%%%%%
Here we consider the same solution $U(x_1)=1-x_1^2$ to (\ref{geometric:eom}) under $n_1=0$ but without any constraint on $f_1,f_2,f_3$ and $c_1,c_2$. In this case (\ref{geometric:susy}) and (\ref{geometric:Bianchi}) are simplified to
\begin{subequations}
	\begin{align}
	e^{C(x_1)}&=-me^{A(x_1)},\\
	m&=k/n=1/g_2=1/g_3,\\
	f(x_1)&=f_0,\\
	g_1(x_1)&=x_1/m,\\
	e^{-3A(x_1)}&=\fft{k}{f_0}+\sum_{i=1}^3\fft{1}{f_i},\\
	\Sigma_{i=1}^3L_i&=0,\\
	M_{ij}&=|\epsilon_{ijk}|L_k,
	\end{align}\label{Betti}
\end{subequations}
The 4-form equation of motion (\ref{geometric:eom}) reduces to the following algebraic constraint 
\begin{equation}
0=e^{4A}(k(f_1f_2+f_2f_3+f_3f_1)+f_0(f_1+f_2+f_3))-f_0f_1f_2f_3(L_1^2+L_2^2+L_1L_2).\label{Betti:eom}
\end{equation}

Under the coordinate transformation $(x,x_1,x_2,x_3)\to(x_1,x_2,x_3,x_4)$ and the following identifications,
\begin{subequations}
	\begin{align}
	m&\to 1,\\
	k&\to 1,\\
	(1/f_0,1/f_1,1/f_2,1/f_3)&\to(l_1,l_2,l_3,l_4),\\ (L_1,L_2,L_3)&\to(2m_{12}e^{2A},2m_{13}e^{2A},2m_{14}e^{2A}),\\
	\Sigma_{i=1}^4x_id\phi_i&\to P,\\
	dx_i\wedge d\phi_i&\to-l_iJ_i,
	\end{align}
\end{subequations}
the solution (\ref{metric},\,\ref{4form}) with (\ref{Betti}) can be rewritten as ($m_{12}=m_{34}$, $m_{13}=m_{24}$, $m_{14}=m_{23}$)
\begin{subequations}
	\begin{align}
	ds^2=&~e^{2A}\left(-r^2dt^2+\fft{dr^2}{r^2}+(d\psi+P)^2\right)+e^{-A}\sum_{i=1}^4\fft{1}{l_i}\left(\fft{dx_i^2}{1-x_i^2}+(1-x_i^2)d\phi_i^2\right),\\
	F_4=&~dt\wedge dr\wedge\fft{(l_2+l_3+l_4)J_1+(l_1+l_3+l_4)J_2+(l_1+l_2+l_4)J_3+(l_1+l_2+l_3)J_4}{l_1+l_2+l_3+l_4}\nn\\&+\Sigma_{i,j=1}^4m_{ij}J^i\wedge J^j,
	\end{align}\label{Betti:rewrite}
\end{subequations}
and the constraint (\ref{Betti:eom}) is also rewritten as
\begin{equation}
l_1l_2+l_1l_3+l_1l_4+l_2l_3+l_2l_4+l_3l_4=2((m_{12})^2+(m_{13})^2+(m_{14}^2)).\label{Betti:eom:rewrite}
\end{equation}
This solution (\ref{Betti:rewrite}) with the constraint (\ref{Betti:eom:rewrite}) is equivalent to the deformed AdS$_2\times\Sigma_2\times Q^{1,1,1}$ solution with two Betti multiplets in section 3.2 of \cite{Donos:2008ug}\footnote{In fact, we need a slight modification: $F_2\to 2F_2$ in (3.21) of \cite{Donos:2008ug} for a perfect match.}.

Let us pause to understand the geometrical basis for the existence of this solution. Given that $b_2(Q^{1,1,1})=2$, that is, the second Betti number is two, we can explicitly construct two linearly independent harmonic 2-forms. Those forms were used in the Ansatz of $F_4$.  We can made this point more explicitly by computing a few physical quantities of this AdS$_2$ solution: the dyonic charges associated to Betti multiplets and the entropy of an AdS$_4$ black hole whose near horizon geometry corresponds to this AdS$_2$ solution. For the definitions of dyonic charges associated to the Betti multiplets, we follow the conventions of \cite{Azzurli:2017kxo}.

We define electric charges associated to the Betti multiplets as
\begin{align}
	Q_i\equiv\int_{C_i}*_{11}F+\fft12 A\wedge F,\qquad C_i\equiv\mathrm{vol}[\Sigma_2]\wedge*_7(\mathrm{vol}[S^2_i]-\mathrm{vol}[S^2_{i+1}]),
\end{align}
for $i=1,2$, where the 7-dimensional Hodge star is defined within the 7-dimensional manifold equipped with the coordinates $\{\psi,x_i,\phi_i\}$. Here we choose $A$ such that $dA=F$ and $\iota_VA=0$ and therefore the 2nd term in the integrand does not contribute. These electric charges are given explicitly as
\begin{align}
	Q_i=-64|\mathfrak{g}-1|\pi^3\Delta m f_0f_1f_2f_3\left(\fft{1}{f_i}-\fft{1}{f_{i+1}}\right)\left(e^{-3A}-\fft{1}{f_i}-\fft{1}{f_{i+1}}\right)\label{E:charge}
\end{align}
for the solution (\ref{Betti:rewrite}), where we have used $\mathrm{Vol}[\Sigma_2]=4\pi|\mathfrak{g}-1|$ and set the period of $\psi$ coordinate as $\Delta$. Note that we have used `$\mathrm{vol}$' for the volume form and `$\mathrm{Vol}$' for the  actual volume of a manifold.

We define magnetic charges associated to the Betti multiplets as
\begin{align}
	P_i\equiv\int_{h_i}F,\qquad h_i\equiv\mathrm{vol}[\Sigma_2]\wedge(\mathrm{vol}[S^2_i]-\mathrm{vol}[S^2_{i+1}]),
\end{align}
for $i=1,2$, which are given explicitly as
\begin{align}
	P_i=16|\mathfrak{g}-1|\pi^2e^{-2A}f_0(f_iL_i-f_{i+1}L_{i+1})\label{M:charge}
\end{align}
for the solution (\ref{Betti:rewrite}).  These charges would correspond, on the field theory side, to baryonic charges as they are related to the topology of $Q^{1,1,1}$.

Finally, considering the solution (\ref{Betti:rewrite}) as a near horizon geometry of an AdS$_4$ black hole, we can compute the Bekenstein-Hawking entropy as
\begin{equation}
	S\equiv\fft{\mathrm{Vol[\text{M}_9]}}{4G_{11}}=\fft{4(kf_1f_2f_3+f_0(f_1f_2+f_2f_3+f_3f_1))|\mathfrak{g}-1|\pi^\fft32|N|^\fft32}{(f_1f_2+f_2f_3+f_3f_1)^\fft32\Delta^\fft12|m|^\fft12},\label{Entropy}
\end{equation}
where the 11-dimensional Newton's constant $G_{11}$ and the flux quantization $N$ are given as
\begin{subequations}
	\begin{align}
	G_{11}&=\fft{(2\pi)^8(L_p^{\scriptscriptstyle{(11)}})^{9}}{16\pi},\\
	N&=\fft{1}{2\pi(L_p^{\scriptscriptstyle{(11)}})^6}\int_{Y_7}*_{11}F+\fft12 A\wedge F=-\fft{m\Delta(f_1f_2+f_2f_3+f_3f_1)}{\pi^3(L_p^{\scriptscriptstyle{(11)}})^6},
	\end{align}
\end{subequations}
in terms of the 11-dimensional Planck length $L_p^{\scriptscriptstyle{(11)}}$. Here we follow the same convention for $A$ that we have used computing electric charges associated to the Betti multiplets.

%%%%%
\subsubsection*{One Betti multiplet}
%%%%%
The solution (\ref{metric},\,\ref{4form}) with (\ref{Betti}) and (\ref{Betti:eom}) has been studied in different conventions for $f_2=f_3$ and $L_2=L_3$, which turns off the charges $Q_2$ and $P_2$ defined in (\ref{E:charge}) and (\ref{M:charge}) and therefore removes the second Betti multiplet. To be specific, under the coordinate transformation $z=1/r$ and the following identifications,
\begin{subequations}
	\begin{align}
	&m\to1,\qquad e^A\to L,\qquad xd\phi\to\mathcal A,\qquad x_2d\phi_2+x_3d\phi_3\to\mathcal A_{\mathcal B},\\
	&e^{-3A}f_0\to u,\qquad e^{-3A}f_1\to v,\qquad -\fft12e^{-2A}f_2L_1\to w,
	\end{align}\label{toABCMZ}
\end{subequations}
the resulting solution can be rewritten as (with $q=1/2$)
\begin{subequations}
	\begin{align}
	ds^2=&~L^2\left(\fft{-dt^2+dz^2}{z^2}+u\left(\fft{dx^2}{1-kx^2}+(1-kx^2)d\phi^2\right)+v\left(\fft{dx_1^2}{1-x_1^2}+(1-x_1^2)d\phi_1^2\right)\right.\nn\\
	&\left.\qquad+\fft{4quv}{-u+v(u-k)}\sum_{i=2}^3\left(\fft{dx_i^2}{1-x_i^2}+(1-x_i^2)d\phi_i^2\right)+(d\psi+2q\mathcal A_{\mathcal B}+k\mathcal A+x_1d\phi_1)^2\right),\\
	F=&~L^3\,\fft{dt\wedge dz}{z^2}\wedge\left((u-k)dx\wedge d\phi+(v-1)dx_1\wedge d\phi_1-2q\fft{u+v(u+k)}{u-v(u-k)}d\mathcal A_{\mathcal B}\right)\nn\\&+wuL^3dx\wedge d\phi\wedge\left(\fft{u-v(u-k)}{u}dx_1\wedge d\phi_1+d\mathcal A_{\mathcal B}\right)\nn\\
	&+wvL^3d\mathcal A_{\mathcal B}\wedge\left(dx_1\wedge d\phi_1+\fft{2u}{u-v(u-k)}d\mathcal A_{\mathcal B}\right),
	\end{align}
\end{subequations}
which is equivalent to the solution (4.22) in \cite{Azzurli:2017kxo} with some modifications on the last two lines in the 4-form. The constraint on $u,v,w$ in (4.23) of \cite{Azzurli:2017kxo} should also be slightly modified to (just a sign flip for $3w^2(u-v(u-k))^2$ in the numerator)
\begin{equation}
	4q^2(-3k^2v^2+2ku(v-1)v+u^2(v-1)(v+3))-3w^2(u-v(u-k))^2=0,
\end{equation}
which is from the 4-form equation of motion (\ref{Betti:eom}) with $q=1/2$.

Here we compute the physical quantities defined above for general cases with two Betti multiplets to compare the results with \cite{Azzurli:2017kxo}.

We start with the case where the additional constraint $L_1=0$ is imposed, which yields a purely electric Betti multiplet with $P_1=0$. In this case, the 4-form equation of motion (\ref{Betti:eom}) yields
\begin{equation}
	f_1=-\fft{f_2(2f_0+kf_2)}{f_0+2kf_2}.
\end{equation}
This implies, for $k=0,1$, $f_1$ must have the opposite sign of $f_2$ if $f_0$ and $f_2$ have the same sign. The positive definite metric (\ref{metric}) requires, however, all $f_0,f_1,f_2,f_3$ to have the same sign. Therefore,  only $k=-1$ yields a solution with positive definite metric. Then the only non-vanishing Betti charge $Q_1$ is given as
\begin{equation}
	Q_1=64|\mathfrak{g}-1|\pi^3\Delta L^6 \fft{uv(u(v-1)-v)(u(v-3)+v)}{(u(v-1)+v)^2},
\end{equation}
under the map (\ref{toABCMZ}). This is the same as (4.28) of \cite{Azzurli:2017kxo} up to sign.

Next, we consider the cases where the additional constraints are given as $f_0=f_2$ for $k=-1$ and $f_1=f_2$ for $k=1$, both of which yield a purely magnetic Betti multiplet with $Q_1=0$. In these cases, the only non-vanishing Betti charge $P_1$ is given as
\begin{equation}
	P_1=\begin{cases}
	-48|\mathfrak{g}-1|\pi^2L^3\fft{w(3w^2+2)(w^2+1)}{3w^2+1} & (k=-1),\\
	-48\pi^2L^3\fft{w(w^2+2)}{w^2-1} & (k=1),\\
	\end{cases}
\end{equation}
under the map (\ref{toABCMZ}). The charge for $k=-1$ does not  match (4.30) of \cite{Azzurli:2017kxo} but the one for $k=1$ matches (4.33). This mismatch is in fact expected from the mismatch in the metric and the 4-form between (\ref{Betti:rewrite}) above and (4.22) of \cite{Azzurli:2017kxo}.

%%%%%
\subsection{New solutions: baryonic and mesonic charges}\label{new:sol}
%%%%%
In the previous subsections we have considered solutions with $n_1=0$ which have been already reported in the literature. These solutions with $n_1=0$ are naturally dual to field theoretic configurations with baryonic charges.  We now turn $n_1$ on, which is designed to match a mesonic charge on the dual field theory. Since this makes the 4-form equation of motion (\ref{geometric:eom}) quite involved, however, it is difficult to find a general, analytic solution to (\ref{geometric:eom}) with a mesonic twist. We therefore take two different approaches: first we focus on the most general \emph{polynomial} solution to (\ref{geometric:eom}) with $n_1\neq0$; then we construct a numerical solution to the same equation of motion.

%%%%%
\subsubsection{A new regular solution in a disconnected branch }\label{poly:sol}
%%%%%
Here we focus on the most general solution to (\ref{geometric:eom}) in the  \emph{polynomial} class. 

Assume that $U(x_1)$ is an arbitrary $p$-th $(p>2)$ degree polynomial function of $x_1$ then we may write $U(x_1)$ explicitly as
\begin{equation}
	U(x_1)=\sum_{i=0}^pa_ix_1^i.\label{poly}
\end{equation}
Substituting (\ref{poly}) into (\ref{geometric:eom}) then simplifies the 4-form equations of motion to
\begin{equation}
	0=-a_p^2(p-1)^2(p+1)(2p-1)f_1^4f_2f_3n_1^4x_1^{2p}+\mathcal O(x_1^{2p-1}),
\end{equation}
which cannot be satisfied unless $a_p=0$. Therefore the most general polynomial solution to the 4-form equation of motion (\ref{geometric:eom}) is at most quadratic since we have demonstrated that $a_p=0$ for $p>3$.

One can check that the most general polynomial solution to (\ref{geometric:eom}) with real coefficients is obtained only when $c_2=0$, and given explicitly as
\begin{equation}
	U(x_1)=\fft{1}{f_1n_1^2}\left(kf(x_1)+\fft{2(f_2+f_3)\pm\sqrt{4(f_2+f_3)^2+3f_2^2f_3^2(c_1^2-4/3f_2f_3)}}{3f_2f_3}f(x_1)^2\right),\label{Sol:poly}
\end{equation}
where $f(x_1)=f_0+f_1\, n_1\, x_1$ is linear in $x_1$. Note that this solution is ill-defined for $n_1=0$.  Implying, therefore,  that the solution we have found lives in a different branch that cannot be smoothly connected to the various solutions we described for $n_1\to 0$.

To the best of our knowledge, the solution (\ref{Sol:poly}) and the corresponding metric and 4-form (\ref{metric},\,\ref{4form}) have not been reported in the literature. We are interested in a compact and regular solution, however, and therefore we still need to investigate if (\ref{Sol:poly}) truly yields a globally well-defined solution. To be more precise, we want a solution with a positive definite metric, finite volume (compact) and which is also singularity-free (regular).  Positive definiteness of the metric (\ref{metric})  requires 
\begin{equation}
	e^{-A(x_1)}f(x_1),\qquad e^{-A(x_1)}f_1U(x_1),\qquad e^{-A(x_1)}f_i~~(i=2,3),\label{positive:def}
\end{equation}
to be positive. The finite volume condition requires the range of $x_1$ to be bounded. Lastly, the singularity-free condition requires that any possible conical singularities should be removed and the final solution without conical singularities must also not have curvature singularities.

First, substituting (\ref{Sol:poly}) into (\ref{Susy:eA}) gives
\begin{equation}
	e^{-A(x_1)}f_i=f_i\left(-\fft{f_2f_3}{f_2+f_3\pm\sqrt{4(f_2+f_3)^2+f_2^2f_3^2(3c_1^2-4/f_2f_3)}}\right)^{-\fft13},\quad(i=2,3)\label{positive:def:1}
\end{equation}
which has to be positive for a positive definite metric. This requires $f_2$ and $f_3$ to have the same sign. For positive (negative) $f_2$ and $f_3$, $e^{-A(x_1)}$ should also be positive (negative) and therefore we need negative[positive] sign in (\ref{positive:def:1}). Since the signs in (\ref{positive:def:1}) are correlated with those in (\ref{Sol:poly}), this means that we must choose negative[positive] sign in (\ref{Sol:poly}) for positive (negative) $f_2$, $f_3$, and $e^{-A(x_1)}$. From here on, we choose positive $e^{-A(x_1)}$ without loss of generality and therefore go with the negative sign in (\ref{Sol:poly}). 

Second, we require $f_1U(x_1)>0$ for a positive definite metric, which yields
\begin{align}
	\begin{cases}
	x\in(x_-,x_+) & c_1^2>\fft{4}{3f_2f_3}, \\
	x\in(-\infty,x_-)\cup(x_+,\infty) & c_1^2<\fft{4}{3f_2f_3},\\
	kf(x_1)>0 & c_1^2=\fft{4}{3f_2f_3},
	\end{cases}
\end{align}
where $x_\pm$ are the zeros of $U(x_1)$. Note that the solution has a finite volume only when $c_1^2>\fft{4}{3f_2f_3}$, so we assume this to be the case. 

Third, we need $f(x_1)>0$ where $x_1\in(x_-,x_+)$ for a positive definite metric since we have chosen $e^{-A(x_1)}>0$. It is equivalent to $f(x_\pm)\geq0$ because $f(x_1)$ is a linear function of $x_1$. Under the constraint $c_1^2>\fft{4}{3f_2f_3}$, this leads to $k>0$ and therefore we must impose $k=1$.

Finally, we set the period of a coordinate $\phi_1$ to remove possible conical singularities at $x_1=x_\pm$. Substituting $x_1=x_\pm\mp r^2$ with a small local coordinate  $r$, we have
\begin{equation}
	\fft{dx_1^2}{U(x_1)}+U(x_1)D\phi_1^2\sim \fft{4}{U'(x_\pm)}\left(d r^2+\fft{U'(x_\pm)^2\, r^2}{4}D\phi_1^2\right),
\end{equation}
where $D\phi_1=d\phi_1+n_1xd\phi$, which determines the period of $\phi_1$ as $4\pi/|U'(x_\pm)|=4\pi|n_1|$.

It is worth mentioning that the solution is also free of curvature singularities. We have computed various curvature invariants including,  $R,R^{ab}R_{ab},R^{abcd}R_{abcd}$ , for particular values of external parameters ($m=f_1=f_2=f_3=1$ and $c_1=2$) and have found that all are bounded constants. 

Now, the solution (\ref{metric}) and (\ref{4form}) satisfying all the conditions discussed above can be rewritten as
\begin{subequations}
	\begin{align}
	ds^2=&~e^{2A}\left(-r^2dt^2+\fft{dr^2}{r^2}+(md\psi+xd\phi+\tilde g_1(\tilde x_1)D\tilde\phi_1+x_2d\phi_2+x_3d\phi_3)^2\right)\nn\\&+e^{-A}\left(\tilde f(\tilde x_1)\left(\fft{dx^2}{1-kx^2}+(1-kx^2)d\phi^2\right)+\tilde f_1\left(\fft{d\tilde x_1^2}{1-\tilde x_1^2}+(1-\tilde x_1^2)D\tilde\phi_1^2\right)\right.\nn\\&\left.\kern4em+\Sigma_{i=2}^3f_i\left(\fft{dx_i^2}{1-x_i^2}+(1-x_i^2)d\phi_i^2\right)\right),\\
	F=&~dt\wedge dr\wedge\left(\fft{3e^{3A}-2\tilde f_1}{2\tilde f_1}(\tilde f(\tilde x_1)dx\wedge d\phi+\tilde f_1d\tilde x_1\wedge D\tilde\phi_1)+\Sigma_{i=2}^3(e^{3A}-f_i)dx_i\wedge d\phi_i\right)\nn\\
	&+dx\wedge d\phi\wedge c_1\tilde f(\tilde x_1)\left(\tilde f_1d\tilde x_1\wedge D\tilde\phi_1-\fft12\Sigma_{i=2}^3f_idx_i\wedge d\phi_i\right)\nn\\
	&-\fft12c_1\tilde f_1d\tilde x_1\wedge D\tilde\phi_1\wedge(\Sigma_{i=2}^3f_idx_i\wedge d\phi_i)+c_1f_2f_3dx_2\wedge d\phi_2\wedge dx_3\wedge d\phi_3,
	\end{align}\label{poly:corresponding:sol}
\end{subequations}
in terms of the new coordinates $(\tilde x_1,\tilde\phi_1)$ introduced as
\begin{equation}
	x_1=\fft{\tilde f(x_1)-f_0}{f_1n_1},\qquad \phi_1=2n_1\tilde\phi_1,
\end{equation}
and the following definitions:
\begin{subequations}
	\begin{align}
	D\tilde\phi_1&=d\tilde\phi_1+\fft12xd\phi,\quad\tilde f(\tilde x_1)=\fft{\tilde f_1}{2}(1+\tilde x_1),\quad
	\tilde g_1(\tilde x_1)=-\fft12(1-3\tilde x_1),\\
	\tilde f_1&=\fft{3f_2f_3}{-2f_2-2f_3+\sqrt{4(f_2+f_3)^2+f_2^2f_3^2(3c_1^2-4/f_2f_3)}},\\
	e^{3A}&=\fft{f_2f_3}{-f_2-f_3+\sqrt{4(f_2+f_3)^2+f_2^2f_3^2(3c_1^2-4/f_2f_3)}}.
	\end{align}
\end{subequations}
Note that the above solution is in fact independent of $n_1$, but still distinguished from the known solutions we have seen in the previous subsections \ref{universal:sol} and \ref{Betti:sol}. Here the ranges of $x$, $\tilde x_1$, $x_2$, and $x_3$ coordinates are $(-1,1)$ and the ranges of $\phi$, $\tilde\phi_1$, $\phi_2$, and $\phi_3$ coordinates are $(0,2\pi)$. We set the range of $\psi$ coordinate to be $(0,\Delta)$. This form of the solution is similar to the expression for the universal twist in which the charge is no longer a free parameter but rather takes a particular value dictated by a general constraint. 

Considering the above solution as a near horizon geometry of an AdS$_4$ black hole, we can compute the Bekenstein-Hawking entropy as
\begin{equation}
	S\equiv\fft{\mathrm{Vol[\text{M}_9]}}{4G_{11}}=\fft{4\sqrt2\tilde f_1^2\pi^\fft32|N|^\fft32}{(f_2f_3\Delta|m|)^\fft12(2\tilde f_1e^{-A}-3e^{2A})^\fft32},
\end{equation}
where the 11-dimensional Newton's constant $G_{11}$ and flux quantization $N$ are given as
\begin{subequations}
	\begin{align}
	G_{11}&=\fft{(2\pi)^8(L_p^{\scriptscriptstyle{(11)}})^{9}}{16\pi},\\
	N&=\fft{1}{2\pi(L_p^{\scriptscriptstyle{(11)}})^6}\int_{\text{M}_7}\star_{11}F+\fft12 A\wedge F=-\fft{mf_2f_3\Delta(2\tilde f_1e^{-3A}-3)}{\pi^3(L_p^{\scriptscriptstyle{(11)}})^6},
	\end{align}
\end{subequations}
in terms of the 11-dimensional Planck length $L_p^{\scriptscriptstyle{(11)}}$. Here M$_7$ denotes the 7-dimensional manifold with the coordinates $\{\psi,\tilde x_1,\tilde\phi_1,x_2,\phi_2,x_3,\phi_3\}$ and we choose $A$ such that $dA=F$ and $\iota_VA=0$. So the 2nd term in the integrand does not contribute.

%%%%%
\subsubsection{A numerical solution with baryonic and mesonic charges}\label{numeric:sol}
%%%%%
The polynomial solution in the previous section \ref{poly:sol} is not valid for $n_1=0$ and does not have a smooth limit as  $n_1\to0$. This obstruction to turning off the would-be mesonic charge $n_1$ indicates that such solution may not represent the solution with a mesonic twist we actually seek\footnote{Furthermore, even for $n_1\neq0$ cases, the corresponding solution (\ref{poly:corresponding:sol}) is in fact independent of the parameter $n_1$, together with the obstruction to turning off $n_1$ which strongly implies that (\ref{poly:corresponding:sol}) is not the solution with a mesonic twist.}. Propelled by this observation, we construct a numerical solution to the 4-form equation of motion (\ref{geometric:eom}) which allows one to smoothly turn off the would-be mesonic charge $n_1$.

Recall that the 4-form equation of motion (\ref{geometric:eom}) is a 4th order non-linear ODE for $U(x_1)$ with the parameters $f_0,f_1,f_2,f_3,c_1,c_2,n_1,k$ which we will refer to as external, where we have substituted $f(x_1)=f_0+f_1n_1x_1$ (\ref{Susy:f}) into (\ref{geometric:eom}). Hence, for given external parameters, we need four initial conditions for $U(x_1)$ to solve (\ref{geometric:eom}) numerically. At this point, the physical constraints for regularity and smoothness  considered in the previous subsection \ref{poly:sol} turn out to be useful in determining those initial conditions.

We start with the two physical constraints: positive definiteness of the metric (\ref{metric}) and  compactness of the global solution. First, the positive definiteness of the metric (\ref{metric}) requires that all of the quantities given in (\ref{positive:def}), namely, $e^{-A(x_1)}f(x_1)$, $e^{-A(x_1)}f_1U(x_1)$, and $e^{-A(x_1)}f_i~(i=2,3)$,  to be positive definite. Then compactness requires that the domain of the coordinate $x_1$ where these quantities are positive must be bounded. If we set $e^{-A(x_1)}$ and $f_1$ to be positive, this implies that we should look for a numerical solution $U(x_1)$ to the ODE (\ref{geometric:eom}) defined on a bounded interval $x_1\in[x_L,x_R]$, which is positive within $x_1\in(x_L,x_R)$ and vanishes at the boundary points $x_1=x_L,x_R$. Note that $f_1$ can be chosen to be positive since it is an external parameter but we have to check if $e^{-A(x_1)}$ is truly positive afterwards.

Now we make the following Ansatz $U_s(x_1)$ around the left boundary point $x_1=x_L$,
\begin{equation}
	U_s(x_1)=\sum_{J=1}^{J_{\rm max}} \fft{u_J}{J!}(x_1-x_L)^J,\label{series:ansatz}
\end{equation}
to solve the ODE (\ref{geometric:eom}) perturbatively with respect to $x_1-x_L$. Determining the coefficients $u_J$'s in terms of the external parameters listed above by substituting (\ref{series:ansatz}) into the ODE (\ref{geometric:eom}), we can decide the initial conditions for actual numerical solutions $U(x_1)$ to the ODE, namely $U(x_L)=0$, $U'(x_L)=u_1$, $U''(x_L)=u_2$, and $U'''(x_L)=u_3$. As mentioned above, the corresponding numerical solution $U(x_1)$ will yield a physical solution only if it vanishes at some finite distance $x_1=x_R>x_L$ as $U(x_R)=0$ and satisfies $U(x_1)>0$ over the domain $x_1\in(x_L,x_R)$. 

Even if we find such a numerical solution, however, it still has to satisfy one more constraint to be a physical solution: it must be singularity-free. Hence the apparent singularity at the boundary point $x_1=x_L,x_R$ where $U(x_L)=U(x_R)=0$ must be a coordinate singularities. This can be achieved if $|U'(x_L)|=|U'(x_R)|$. To be specific, consider the following series expansion of the 2D metric around the boundary points
\begin{align}
	\fft{dx_1^2}{U(x_1)}+U(x_1)D\phi_1^2~\sim~
	\fft{4\alpha}{U'(x_b)}\left(dr^2+\fft14U'(x_b)^2r^2D\phi_1^2\right),
\end{align}
where $x_1-x_b=\alpha r^2$ is small and $x_b$ stands for $x_L,x_R$. This expansion shows that the apparent singularities both at $x_1=x_L$ and $x_1=x_R$, or equivalently at $r=0$, become coordinate singularities and it is possible to avoid conical singularities by choosing  the period of a coordinate $\phi$ is chosen as $4\pi/|U'(x_b)|$. 

From here on, we set $U'(x_L)=u_1=2$ as one of the initial conditions for a numerical solution and then the corresponding period of a coordinate $\phi$ would be $2\pi$. One remaining condition for a numerical solution $U(x_1)$ over the domain $[x_L,x_R]$ to be physical is then $U'(x_R)=-2$. Note that we can exclude $U'(x_R)=+2$ since $U(x_1)$ is positive within $(x_L,x_R)$. In order to satisfy such a condition, we want to leave at least one tunable parameter in solving the ODE (\ref{geometric:eom}) numerically. Since $u_1$ is fixed to $u_1=2$, we choose $u_2$ as our tunable parameter then the higher order coefficients $u_J~(J\geq3)$ in the series Ansatz (\ref{series:ansatz}) satisfying the ODE (\ref{geometric:eom}) perturbatively with respect to $x_1-x_L$ would be fixed in terms of $u_2$ and the external parameters.

\textbf{Example 1}

\begin{figure}[htb!]
	\centering
	\includegraphics[width=.4\linewidth]{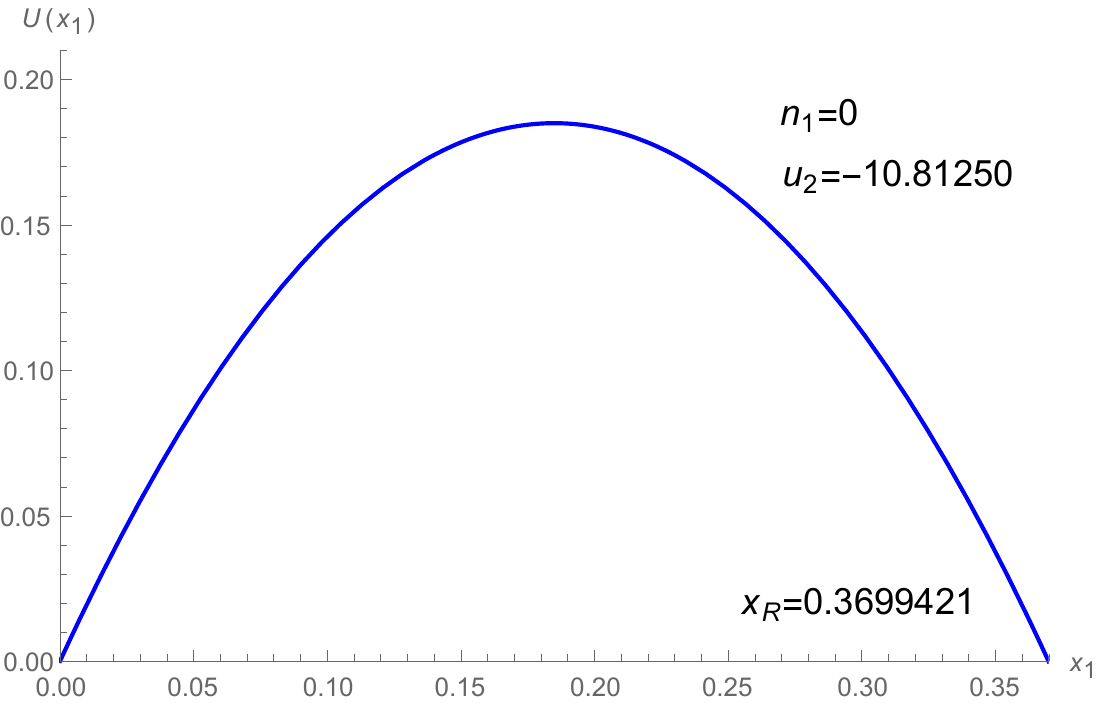}
	\includegraphics[width=.4\linewidth]{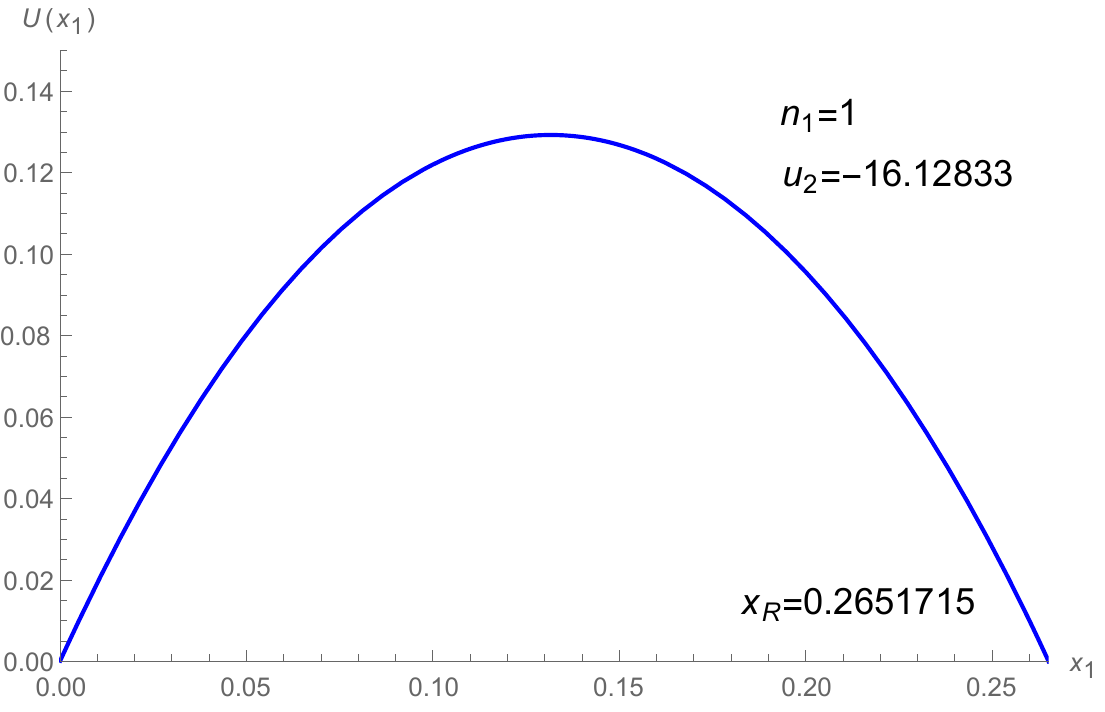}\\[1em]
	\includegraphics[width=.4\linewidth]{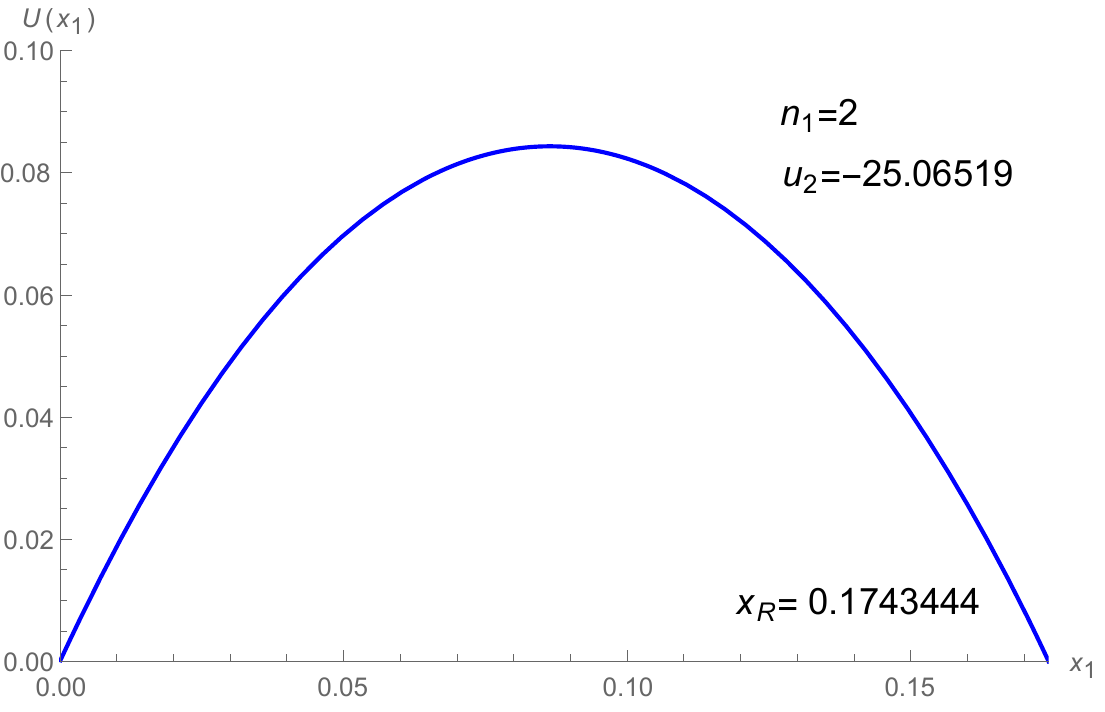}
	\includegraphics[width=.4\linewidth]{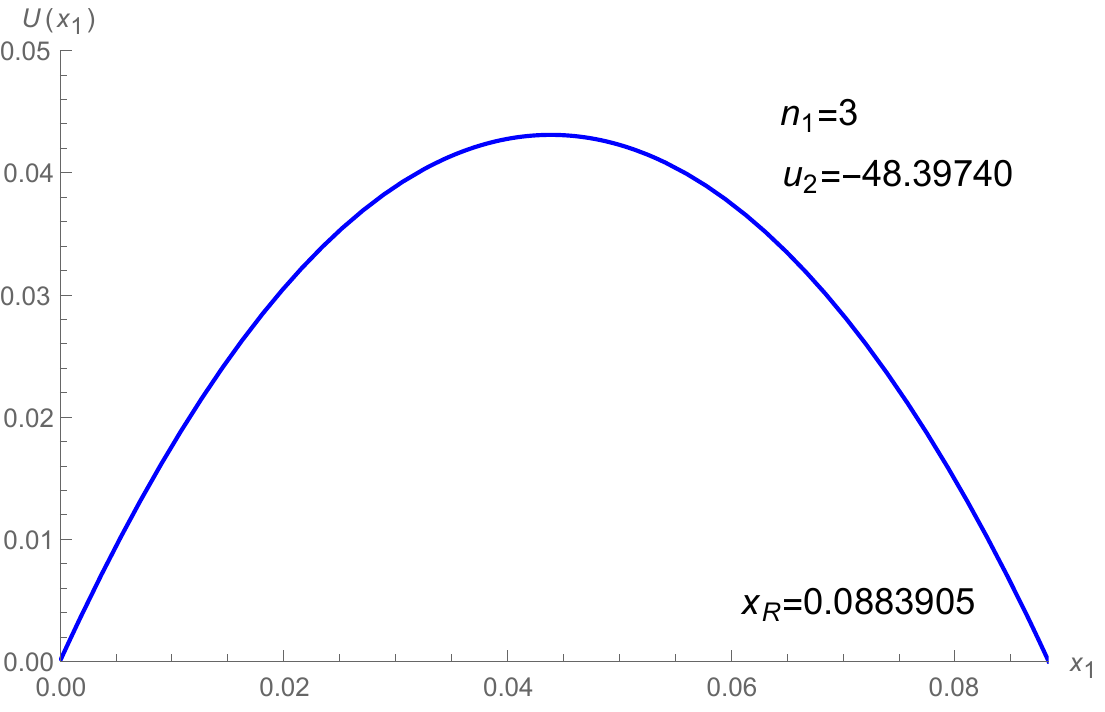}
	\caption{$(f_0,f_1,f_2,f_3,c_1,c_2,k)=(2,1,1,2,4,1,1)$ and $u_1=2$}
	\label{EX1}
\end{figure}

Based on the above setup, we find numerical solutions. Figure \ref{EX1}  shows one example with the external parameters $(f_0,f_1,f_2,f_3,c_1,c_2,k)=(2,1,1,2,4,1,1)$ and $u_1=2$ for three different integers $n_1=0,1,2,3$, where we choose the left boundary point as $x_L=0$ for convenience. We will now show that the solutions satisfy all the physical constraints we have imposed. 

%%%%
%%%%

\begin{figure}[htb!]
	\centering
	\includegraphics[width=.4\linewidth]{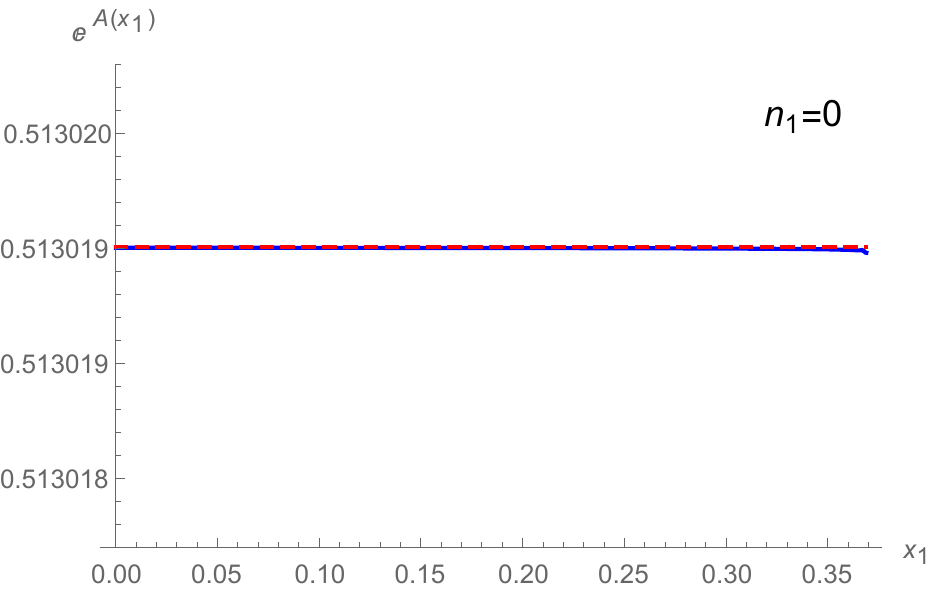}
	\includegraphics[width=.4\linewidth]{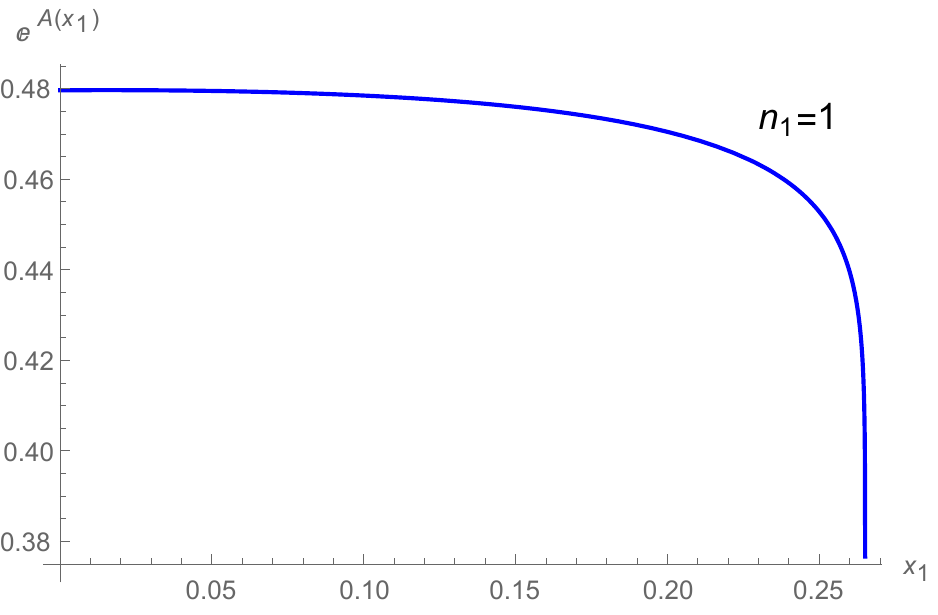}\\[1em]
	\includegraphics[width=.4\linewidth]{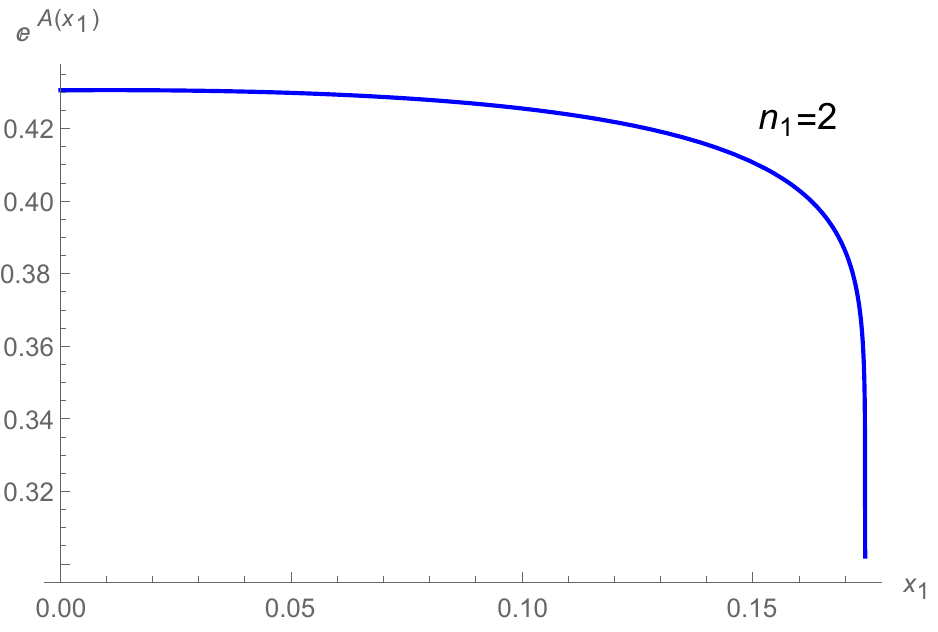}
	\includegraphics[width=.4\linewidth]{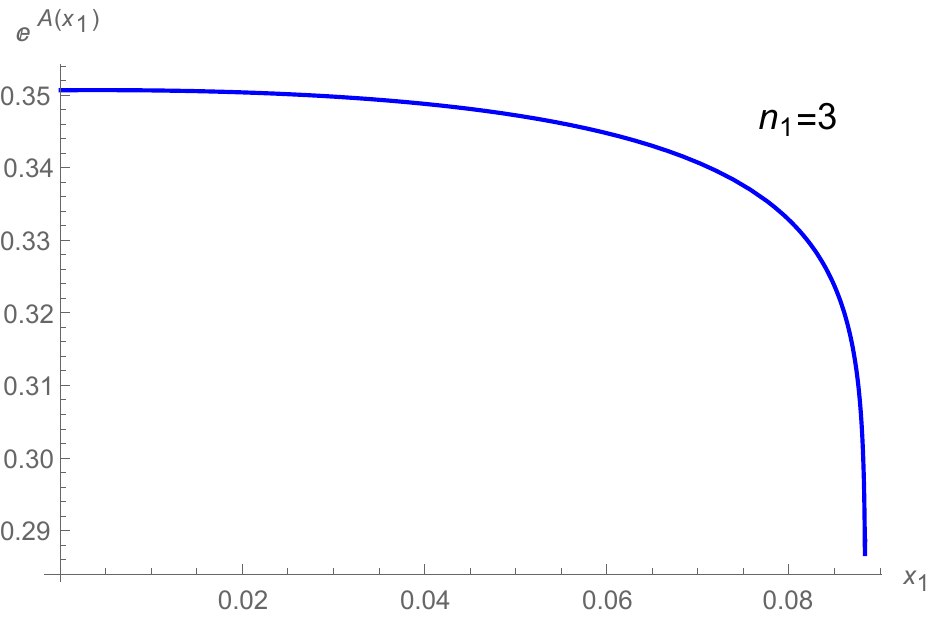}
	\caption{$(f_0,f_1,f_2,f_3,c_1,c_2,k)=(2,1,1,2,4,1,1)$ and $u_1=2$: Note that, in the upper-left panel for $n_1=0$, the numerical plot (blue) matches the constant value of $e^{A(x_1)}$ from the exact quadratic polynomial solution $U(x_1)=-\fft{173}{32}x_1(x_1-x_R)$ (red, dashed).}
	\label{EX1:eA}
\end{figure}

First, $e^{-A(x_1)}$ is positive definite within the domain $x_1\in(x_L=0,x_R)$ for each case as can be seen in Figure \ref{EX1:eA}. Since the numerical solution $U(x_1)$, the linear function $f(x_1)$, and $f_1,f_2$ and $f_3$ are also all positive definite within the same domain, all the quantities in (\ref{positive:def}) are positive definite and therefore we have a positive-definite metric (\ref{metric}). Second, the domain of the $x_1$ coordinate is bounded so the corresponding global solution is compact. Third, $U'(x_R)=-2$ is satisfied for each $n_1$ with the chosen tunable parameters listed in Figure \ref{EX1} and therefore we have $|U'(x_L)|=|U'(x_R)|=2$. This guarantees that the conical singularities at $x_1=x_L,x_R$ can be removed provided that the period of the  coordinate $\phi_1$ is chosen as $2\pi$.

Furthermore, these numerical solutions are continuously deformed under $n_1\to0$ as demonstrated in the Figure \ref{EX1}. Such a smooth behavior makes these numerical solutions perfect candidates for mesonic twisted solutions compared to the polynomial solutions studied in section \ref{poly:sol}.

%%%%
%%%%

\textbf{Example 2}

\begin{figure}[htb!]
	\centering
	\includegraphics[width=.4\linewidth]{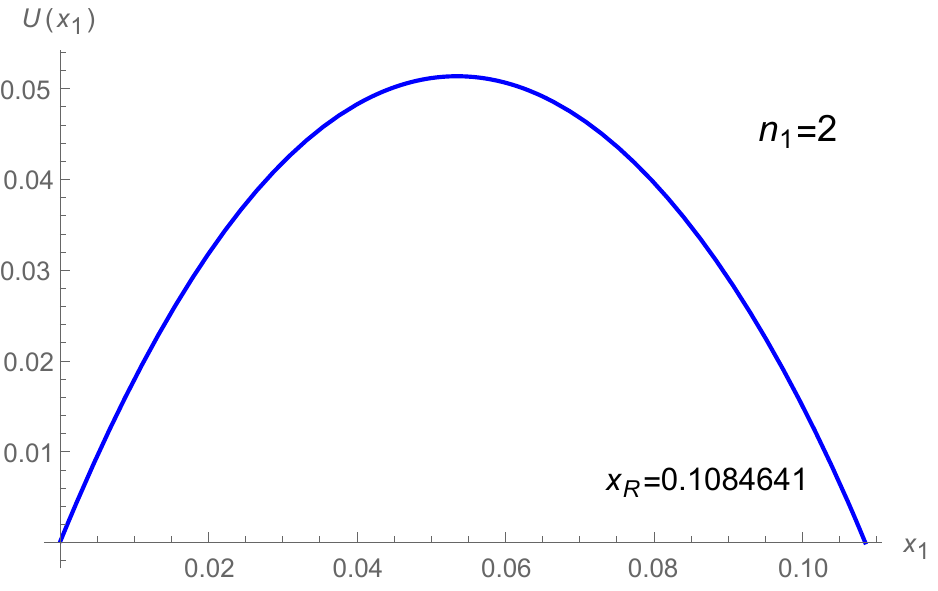}
	\includegraphics[width=.4\linewidth]{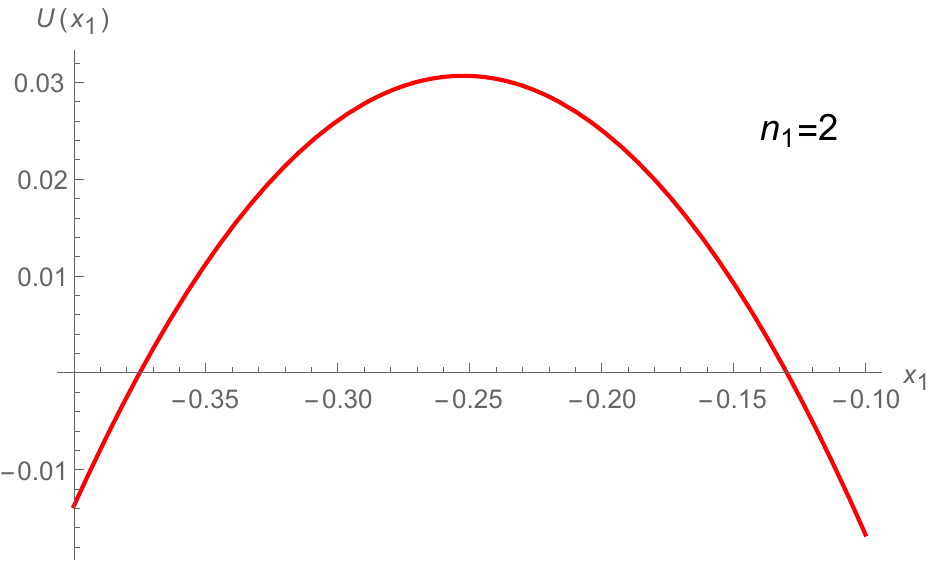}
	\caption{$(f_0,f_1,f_2,f_3,c_1,c_2,k)=(\fft32,2,3,1,3,0,1)$ and $u_2=-21.32431$: the numerical solution(LHS) is distinguished from the polynomial solution(RHS).}
	\label{EX2}
\end{figure}

We find another numerical solution with different external parameters $(f_0,f_1,f_2,f_3,c_1,c_2,k)=(\fft32,2,3,1,3,0,1)$, $u_1=2$, and $n_1=2$. The tunable parameter is chosen as $u_2=-21.32431$ for $U'(x_R)=-2$ as in Example 1 and the corresponding numerical solution is plotted in the left hand side of Figure \ref{EX2}. This set of external parameters satisfies the constraints for a quadratic polynomial solution studied in \ref{poly:sol} to exist, namely $c_2=0$, $c_1^2>\fft{4}{3f_2f_3}$, and $k=1$. We have plotted the corresponding polynomial solution (\ref{Sol:poly}) with the negative sign in the right hand side of Figure \ref{EX2}. Figure \ref{EX2} makes it explicit that the numerical solution obtained in this section are different from the polynomial solutions in a disconnected branch construction in section \ref{poly:sol}.
%%%%
%%%%

%%%%%%%%%%%%%%%%%%%%%%%%%%%%%%%%%%%%%%%%%%%%%%

%%%%%%%%%%%%%%%%%%%%%%%%%%%%%%%%%%%%%%%%%%%%%%%

%%%%%%%%%%%%%%%%%%%%%%%%%%%%%%%%%%%%%%%%%%%%%%%
\section{Conclusions}

In this manuscript we have conducted a partial classification of AdS$_2$ solutions in 11d supergravity.  In the case where the M$_9$ manifold admits an SU(4)-structure the differential and algebraic conditions that the background needs to satisfy are given in \eqref{eq:diffcond1}-\eqref{eq:algebcond3}. Let us briefly summarize the situation. Since an SU(4)-structure is canonically eight-dimensional, we find that the relevant data consists of a real 1-form $V$ strictly orthogonal to the SU(4)-structure in a way that generates M$_9$ metrically (see \eqref{eq:metric}). The $SU(4)$-structure itself is given by a $(1,1)$ form $J$ and a $(4,0)$ form $\Omega$.  The 4-form flux of 11d supergravity is parametrixed in terms of a  2-form and a 4-form $(G_2,G_4)$, see \eqref{eq:AdS2decomp}. Thus, the necessary and sufficient conditions \eqref{eq:diffcond1}-\eqref{eq:algebcond3} are given in terms of $(V, J,\Omega)$ and the fluxes  $(G_2,G_4)$.

From the classification point of view the truly new results in this manuscript pertain to the  necessary and sufficient conditions for ${\cal N}=(1,0)$ supersymmetry \eqref{eq:diffcond1}-\eqref{eq:algebcond3}.  Our result refines the case of ${\cal N}=(2,0)$ susersymmetry  which has been  discussed previously in the literature albeit in a peculiar way through transgression \cite{Donos:2008ug}.  Roughly, descending to the ${\cal N}=(2,0)$ case is equivalent to assuming that M$_9$ is a U(1) fibration over the M$_8$ base that is independent of U(1) coordinate, that is, the one-form $V$ in the general case becomes the one-form dual to the U(1) direction $\partial_\psi$.

It is worth pointing out that our approach, given its direct connection to \cite{Gauntlett:2002fz}, provides an embedding of a class of black-hole near horizons into a context general enough to describe the entire black-hole. Finding the full, interpolating, black holes remains an open problem but our work could easily become a first step in that direction. Having such full interpolating solutions from the near horizon to the asymptotically AdS$_4$  region would help clarify various aspects. For example, by evaluating its on-shell action one could  potentially clarify the ${\cal I}$-extremizaiton procedure \cite{Gauntlett:2019roi,Hosseini:2019ddy,Kim:2019umc} in terms of an attractor mechanism in the bulk extending on previous related work along the lines of \cite{Cabo-Bizet:2017xdr,Halmagyi:2017hmw,BenettiGenolini:2019jdz}. 

When specialized to the case of $Q^{1,1,1}$  our discussion provides a conceptual home for various solutions known in the literature and it allows us to present  new solutions.  We considered a series of generalizations culminating in a numerical analysis providing evidence for the existence of a solution with nontrivial baryonic as well as mesonic charges. We also found a peculiar solution which is disconnected from the $n_1=0$ branch, it would be opportune to track all the potential branches in the future.

It would be quite interesting to extend the classification of supersymmetric AdS$_2$ spaces allowing for more general M$_9$ spaces, that is, beyond the SU(4)-structure case. More physically, in this manuscript we elucidated the situations when the background can ultimately  be understood as deformations of AdS$_4\times $SE$_7$, that is, as backgrounds ultimately originating from M2 branes which naturally admit SU(4)-structure as we have discussed.  It is quite relevant  to extend our analysis in more detail to the  class of supersymmetric solutions with AdS$_2$ factors pertaining to  those arising from wrapped M5 branes. A prototypical class pertains to solutions where the M5 branes wrap a hyperbolic 3-manifold for which the dual field theory solution is well known (see, for example \cite{Gang:2014ema,Gang:2018hjd,Gang:2019uay}). In the asymptotic AdS$_4$ regions the seven dimensional manifold is no longer a U(1) bundle over a Kahler-Einstein 6d manifold as in the case of M2 branes, rather it is a $S^4$ fibration over a hyperbolic 3-manifold  \cite{Gang:2014ema,Bah:2014dsa}.  It is reasonable to expect that the AdS$_2$ classification of such solutions goes beyond the SU(4)-structure case treated here.  Finally, let us point out another class that does not fit in our SU(4)-structure classificatory approach and that would be interesting to tackle - AdS$_3$ solutions in M-theory. Recall that AdS$_3$ can always be written as an AdS$_2$ foliation, therefore, a complete classification of AdS$_2$ solutions must include all AdS$_3$ solutions. However, in appendix \ref{App:NoGoAdS3} we proved the absence of AdS$_3$ solutions within the SU(4)-structure implying that to capture bubbling solutions \cite{Lin:2004nb,Lunin:2007ab,DHoker:2008lup} we need to generalize our work.  We hope to return to these topics in the future.

%%%%%%%%%%%%%%%%%%%%%%%%%%%%%%%%%%%%%%%%%%%%%%%%%%
\section*{Acknowledgments}
We are thankful to Ibou Bah, Francesco Benini,  Marina David, Morteza Hosseini, Nakwoo Kim, Jim Liu, Anton Nedelin, Jun Nian and  Alberto Zaffaroni. We are particularly grateful to Chris Couzens for various in-depth discussions and to Alessandro Tomasiello for important comments. NTM and LPZ would like to express a special thanks to the Mainz Institute for Theoretical Physics (MITP) of the Cluster of Excellence PRISMA+ (Project ID 39083149) for its hospitality and support. JH and LPZ are supported in part by the U.S. Department of Energy under grant DE-SC0007859.  NTM is funded by the Italian Ministry of Education, Universities and Research under the Prin project ``Non Perturbative Aspects of Gauge Theories and Strings'' (2015MP2CX4) and INFN.

\appendix

%%%%%%%%%%%%%%%%%%%%%%%%%%%%%%%%%%%%%%%%%%%%%%%%%%%%%%%
\section{There are no AdS$_2$ solutions with Spin(7)-structure}\label{App:NoGoSpin}
In this Appendix we prove that there are no AdS$_2$ solutions in M-theory with internal space supporting a Spin(7)-structure globally.\\
~\\
Similar to an SU(4)-structure, a Spin(7)-structure in 9 dimensions can be defined in terms of two spinors, that are chiral when viewed in 8 dimensions. This time however these spinors should be equal\footnote{Strictly speaking they need only be parallel.  However the condition that $\chi=\chi_1+ i \chi_2$ should be unit norm means that $\chi= e^{i \alpha}\chi_0$, for $\alpha$ a phase.  This phase can then be set to zero with a frame rotation, so we loose no generality by assuming $\chi_1=\chi_2$.}.  The canonical dimension of a Spin(7)-structure is 8, so  M$_9$ will decompose as
\beq
ds^2(\text{M}_9)= ds^2(\text{M}_8) +U^2,
\eeq 
with $\text{M}_8$ the manifold supporting the Spin(7)-structure and with $U$ a real 1-form that sits orthogonal to it.\\
~\\
We begin with spinor Ansatz
\beq\label{eq:aspinor}
\epsilon= (\zeta_++\zeta_-)\otimes \chi, 
\eeq 
for $\zeta_{\pm}$ and $\chi$ Majorana spinors on AdS$_2$ and M$_9$ respectively - $\pm$ labels 2d chirality as elsewhere. 
Let us without loss of generality fix the 8 dimensional chirality via the projections
\beq
\gamma_{1...8}\chi=\chi.
\eeq
A Spin(7) structure in 9 dimensions is defined in terms of the 1-form $U$ and a real 4-form $\Psi$ with components given by
\beq
U_a=\chi^{\dag}\gamma_a\chi,~~~~~ \Psi_{a_1..a_4}= \chi^{\dag}\gamma_{a_1..a_4}\chi
\eeq
where $\star_9 \Psi= U\wedge \Psi$. Using these definitions it is a simple matter to establish that the 11 dimensional supersymetric forms are given by 
\begin{subequations}
\begin{align}
K&=- r e^{A}(e^0- e^r),\\[2mm]
\Omega&=K\wedge U,\\[2mm]
\Sigma&=K\wedge\Psi.
\end{align}
\end{subequations}
Where $e^0,e^r$ are the vielbein on warped AdS$_2$ given in \eqref{eq:warpedads2}. As such we now have $||K||^2=0$, so the Killing vector is null rather than time-like.  However if we now attempt to solve the supersymmetry conditions $(K,\Omega,\Sigma)$ should obey, we find that we cannot. Specifically consider \eqref{eq:11dsusy2}, this gives rise to
\beq
(r^2 dt-dr)\wedge(e^{2A}G_2+ d(e^{2A} U))+ 2 r e^{2A}U\wedge dt\wedge dr=0,
\eeq
but the terms in this sum must vanish by themselves, and the second cannot be solved for a non zero spinor - hence there are no such solutions.

%%%%%%%%%%%%%%%%%%%%%%%%%%%%%%%%%%%%%%%%%%%%%%%%%%%%%
\section{No AdS$_3$ solutions within SU(4)-structure AdS$_2$ class}\label{App:NoGoAdS3}
In this section we shall demonstrate that the class of AdS$_2$ solutions in section \ref{eq:neq1ads2} is not exhaustive. We shall do so by proving that it contains no AdS$_3$ solutions which are known to exist in M-theory \cite{Martelli:2003ki}.\\
~\\
 AdS$_3$  can be expressed as a foliation of AdS$_{2}$ over an interval as
\beq
ds^2(\text{AdS}_3)=\cosh^2 x ds^2(\text{AdS}_{2})+ dx^2. 
\eeq
As such a complete classification of AdS$_2$ solutions should contain all AdS$_3$ solutions as well. To find such solutions within section \ref{eq:neq1ads2} , we must decompose the metric such that
\beq
e^{2A}ds^2(\text{AdS}_2)+ ds^2(\text{M}_8)+ V^2= e^{2A_3}\bigg[\cosh^2 xds^2(\text{AdS}_2)+ dx^2\bigg]+ ds^2(\tilde{\text{M}}_8),
\eeq
and similarly for the fluxes. To achieve this we must clearly fix
\beq\label{eq:A21}
e^{2A}= e^{2A_3}\,\cosh^2x .
\eeq
In general the foliation direction $dx$ can lie partially along $V$ and partially along $\text{M}_8$, which supports the SU(4)-structure, as such we should decompose
\begin{align}\label{eq:A22}
V&= \cos\zeta \,e^{A_3}dx+ \sin \zeta  \,k_1,\\[2mm]
J&= (\cos\zeta \, k_1- \sin \zeta  \, e^{A_3}dx)\wedge k_2+ J_3,\\[2mm]
\Omega&=(\cos\zeta  \,k_1- \sin \zeta  \, e^{A_3}dx+ i k_2)\wedge \Omega_3,
\end{align}
where $J_3,\Omega_3$ are the (1,1) and (3,0) forms defining an SU(3) structure in 6 dimensions, and $k_1,k_2$ are two real 1-forms that together with the other six dimensions span $\tilde{\text{M}}_8$. The angle $\zeta$ is point dependent on $\tilde{\text{M}}_8$ and defines the alignment of $dx$. 
The above relations are analogous to a set  introduced in \cite{Gauntlett:2004zh} while discussing supersymmetric AdS$_5$ solutions in M-theory.
\\
~\\
The decomposition \eqref{eq:A21}-\eqref{eq:A22} is sufficient to ensure an AdS$_3$ factor without loss of generality, provided the flux also respects the AdS$_3$ isometries.    Unfortunately though, the possibility of AdS$_3$ dies as soon as one considers the supersymmety constraint \eqref{eq:diffcond1}, which decomposes as
\beq
d(e^A J)= \cosh x d\big(e^{A_3}(J_3+ \cos\zeta k_1\wedge k_2)\big)+ \cosh x dx\wedge d(e^{2A_3} \sin\zeta k_2)+ e^{A_3}\sinh x dx\wedge (J_3+ \cos\zeta k_1\wedge k_2).  \nn
\eeq
The issue is the final term in this expression  which requires that $J_3+ \cos\zeta k_1\wedge k_2=0$, since $J_3,k_1,k_2$ are by definition non zero and mutually orthogonal there is no way to solve this.\\ 
~\\
We have thus shown that there are no AdS$_3$ solutions contained in the class of section \ref{eq:neq1ads2}, and as such this class is clearly not the whole story for $\mathcal{N}=1$ AdS$_2$ in M-theory. Recall that there is a well-understood set of bubbling solutions of the form AdS$_3\times S^3\times S^3\times \Sigma_g$ \cite{Lin:2004nb,Lunin:2007ab,DHoker:2008lup}  that plays an important role in the AdS/CFT correspondence. There is also AdS$_2$ bubbling \cite{Bena:2018bbd,Li:2018omr}.

Let us conclude this appendix by recalling that in \cite{Kim:2006qu} a solution containing AdS$_2$ was found exploiting a foliation  of AdS$_4$ in the standard Freund-Rubin AdS$_4\times $SE$_7$ solution.  Note that realizing AdS$_4$ from AdS$_2$ requires one to allow $J_3$ and $\Omega$ to depend on $x$ above.  We assume they do not since we are  interested in AdS$_3$ with compact internal space, that is,  AdS$_3$ that is not part of AdS$_4$ or some higher AdS factor.

%%%%%%%%%%%%%%%%%%%%%%%%%%%%%%%%%%%%%%%%%%%%%%%%
\section{Killing spinor approach}\label{App:KillingSpinor}
%%%%%
The metric ansatz and the corresponding coframe are the same as (\ref{metric}) and (\ref{coframe}). The 4-form ansatz is also the same as (\ref{4form}) with $G_2$ chosen explicitly as
\begin{equation}
	G_2=H(x_1)e^4\wedge e^5+\Sigma_{i=1}^3J_i(x_1)e^{2i+4}\wedge e^{2i+5}+K_1(x_1)e^3\wedge e^6.
\end{equation}

Now we solve the Killing spinor equations, the 4-form Bianchi identity, and the 4-form equations of motion following the conventions of \cite{Gauntlett:2002fz} (Einstein equations will automatically follows):
\begin{align}
	0&=\nabla_\mu\epsilon+\fft{1}{288}\left(\Gamma_\mu{}^{\nu_1\nu_2\nu_3\nu_4}-8\delta_\mu{}^{\nu_1}\Gamma^{\nu_2\nu_3\nu_4}\right)F_{\nu_1\nu_2\nu_3\nu_4}\epsilon,\label{Killing}\\
	0&=dF,\label{Bianchi}\\
	0&=d*F+\fft12F\wedge F,\label{EOM}\\
	0&=R_{\mu\nu}-\fft{1}{12}\left(F_{\mu\nu_1\nu_2\nu_3}F_{\nu}{}^{\nu_1\nu_2\nu_3}-\fft{1}{12}g_{\mu\nu}F^2\right).
\end{align}

First, to solve the Killing spinor equations, we choose the following projections,
\begin{equation}
	\Gamma^1\epsilon=\alpha\epsilon,\quad \Gamma^{23}\epsilon=\beta\epsilon,\quad\Gamma^{45}\epsilon=\gamma\epsilon,\quad\Gamma^{2i+4,2i+5}=\delta_i\epsilon~(i=1,2,3),
\end{equation}
where $\alpha,\beta,\gamma,\delta_i\in\{\pm i\}$. Note that the above projections are given with respect to the coframe index. Under these projections, the Killing spinor equations (\ref{Killing}) yield the differential conditions,
\begin{subequations}
\begin{align}
	0&=\partial_r\epsilon-\fft{1}{2r}\epsilon,\\
	0&=(n\partial_\psi+n_1\partial_{\phi_1})\epsilon-\fft12k\gamma\epsilon,\\
	0&=\partial_{x_1}\epsilon-\fft12A'(x_1)\epsilon,\\
	0&=\partial_\psi\epsilon-\fft{\delta_2}{2g_2}\epsilon=\partial_\psi\epsilon-\fft{\delta_3}{2g_3}\epsilon,
\end{align}\label{killing:susy:1}
\end{subequations}
where $\partial_{\mu}\epsilon=0$ for $\mu=t,\theta,\phi,x_i,\phi_i~(i=2,3)$. We can also derive the algebraic conditions,
\begin{subequations}
\begin{align}
	e^{C(x_1)}&=-me^{A(x_1)},\\
	m\beta&=\delta_2/g_2=\delta_3/g_3,\\
	H(x_1)&=-\alpha\beta\gamma\,e^{2A(x_1)}m\left(\delta_1\fft{g'(x_1)}{f_1}+\delta_2\fft{g_2}{f_2}+\delta_3\fft{g_3}{f_3}\right),\\
	J_1(x_1)&=\alpha\left(-\beta\,e^{2A(x_1)}m\fft{g'(x_1)}{f_1}+\delta_1 e^{-A(x_1)}\right),\\
	J_i(x_1)&=\alpha\left(-\beta\,e^{2A(x_1)}m\fft{g_i}{f_i}+\delta_i\,e^{-A(x_1)}\right)\quad(i=2,3),\\
	K_1(x_1)&=-3\alpha\beta\,e^{A(x_1)/2}\sqrt{U(x_1)}A'(x_1),\\
	f(x_1)&=f_0-n_1\gamma\delta_1 x_1,\\
	e^{3A(x_1)}&=\fft{2f_1f_2f_3f(x_1)}{2f_1f_2f_3(k+\gamma\delta_1n_1U'(x_1))+f(x_1)(2f_1(f_2+f_3)-f_2f_3U''(x_1))},\\
	g_1(x_1)&=-\fft{k}{mn_1}\beta\gamma-\fft{n}{n_1}+\fft{1}{2m}\beta\left(\fft{n_1U(x_1)}{f(x_1)}\gamma+U'(x_1)\delta_1\right),\\
	M_{ij}(x_1)&=\gamma\delta_i\delta_j\delta_k|\epsilon_{ijk}|L_k(x_1),\\
	0&=\Sigma_{i=1}^3\delta_iL_i(x_1).
\end{align}\label{killing:susy:2}
\end{subequations}
where $m$ is a constant. 

Provided that the Killing spinor equations (\ref{killing:susy:1}) and (\ref{killing:susy:2}) are satisfied, the 4-form Bianchi identity (\ref{Bianchi}) yields
\begin{subequations}
\begin{align}
	L_1(x_1)&=c_1e^{2A(x_1)},\\
	L_2(x_1)&=\left(-\fft{\delta_1c_1}{2\delta_2}+\fft{c_2}{f(x_1)^2}\right)e^{2A(x_1)}.
\end{align}\label{killing:bianchi}
\end{subequations}

Finally, the 4-form equations of motion (\ref{EOM}) yields a constraint on the projections and the 4th order ODE for $U(x_1)$,
\begin{align}
	\alpha=&~\beta\gamma\delta_1\delta_2\delta_3,\\
	0=&~f_1^2f_2f_3(4c_2^2+f_1^2n_1^4U(x_1)^2)+2\gamma\delta_1f(x_1)f_1^3f_2f_3n_1^3\left(U(x_1)U'(x_1)\right)\nn\\&+f(x_1)^2f_1^2f_2f_3n_1^2\left(U'(x_1)^2+U(x_1)U''(x_1)\right)\nn\\&+f(x_1)^3f_1\left(-4f_1(f_2+f_3)(k+\gamma\delta_1n_1U'(x_1))+f_2f_3(2kU''(x_1)+3\gamma\delta_1n_1(U(x_1)U''(x_1))')\right)\nn\\&+f(x_1)^4\left(-4f_1^2+3f_1^2f_2f_3c_1^2+2f_1(f_2+f_3)U''(x_1)-f_2f_3(U(x_1)U'''(x_1))'\right).\label{killing:eom}
\end{align}

This expression coincides, of course, with \eqref{geometric:eom} obtained in the main text using the geometric SU(4) structure conditions and the 4-form equation of motion.

\bibliographystyle{JHEP}
\bibliography{Localization}
%\nocite{*}

\end{document}